\documentclass{ar-1col}
\usepackage{url}
\usepackage[numbers]{natbib}
\usepackage{multirow}
\usepackage{graphicx}
\usepackage{etoolbox}
\usepackage{color}
\usepackage{bm}
\usepackage{amssymb,amsfonts,amsmath}
\usepackage{setspace}
\usepackage{subcaption} 
\usepackage{wrapfig}

\newif\iffigures
\figurestrue 

\newif\ifcomments
\commentstrue 

\ifcomments
\newcommand{\EF}[1]{{\color{magenta}EF: #1 \color{black}}}
\newcommand{\LF}[1]{{\color{blue}LF: #1 \color{black}}}
\newcommand{\AR}[1]{{\color{darkred}AR: #1 \color{black}}}
\newcommand{\AK}[1]{{\color{darkgreen}AK: #1 \color{black}}}
\else
 \newcommand{\EF}[1]{}
 \newcommand{\LF}[1]{}
 \newcommand{\AR}[1]{}
 \newcommand{\AK}[1]{}
\fi

\newcommand{\Lopt}{{\ell}_{\text{opt}}} 

\newcommand{\Seve}{{S}_{\text{eve}}}
\newcommand{\Lanh}{{L}_\text{anh}} 
\newcommand{\Lc}{{{L}_{\text{c}}}}

\newcommand{\dep}{_\text{dep}}

\newcommand{\zetadep}{\zeta_{\text{dep}}}

\newcommand{\zetaeq}{\zeta_{\text{eq}}}

\newcommand{\taudep}{\tau_{\text{dep}}}
\newcommand{\taueq}{\tau_{\text{eq}}}

\newcommand{\nudep}{\nu_{\text{dep}}}

\newcommand{\nueq}{\nu_{\text{eq}}}

\newcommand{\zetadepanh}{\zeta^{\text{qKPZ}}_{\text{dep}}}
\newcommand{\taudepanh}{\tau^{\text{qKPZ}}_{\text{dep}}}

\definecolor{darkblue}{rgb}{0,0,0.6}
\definecolor{darkred}{rgb}{0.6,0,0}
\definecolor{darkgreen}{rgb}{0,0.6,0}

\jname{Xxxx. Xxx. Xxx. Xxx.}
\jvol{AA}
\jyear{YYYY}
\doi{10.1146/((please add article doi))}

\begin{document}

\markboth{Ferrero et al.}{Creep motion}

\title{Creep motion of elastic interfaces driven in a disordered landscape} 

\author{E. E. Ferrero,$^1$ L. Foini,$^2$ T. Giamarchi,$^3$ A. B. Kolton,$^4$ and A. Rosso$^5$
\affil{$^1$Instituto de Nanociencia y Nanotecnolog\'{\i}a,
Centro At\'omico Bariloche, CNEA--CONICET, R8402AGP San Carlos de Bariloche, R\'{\i}o Negro, Argentina}
\affil{$^2$IPhT, CNRS, CEA, Universit\'{e} Paris-Saclay, 91191 Gif-sur-Yvette, France}
\affil{$^3$Department of Quantum Matter Physics, University of Geneva,
24 Quai Ernest-Ansermet, CH-1211 Geneva, Switzerland}
\affil{$^4$ Instituto Balseiro, Centro At\'omico Bariloche, CNEA--CONICET--UNCUYO, R8402AGP San Carlos de Bariloche, R\'{\i}o Negro, Argentina}
\affil{$^5$LPTMS, CNRS, Univ. Paris-Sud, Universit\'{e} Paris-Saclay, 91405 Orsay, France}
}

\begin{abstract}
The thermally activated creep motion of an elastic interface weakly driven
on a disordered landscape is one of the best examples of glassy universal dynamics.
Its understanding has evolved over the last 30 years thanks to a fruitful interplay
between elegant scaling arguments, sophisticated analytical calculations, efficient
optimization algorithms and creative experiments.
In this article, starting from the pioneer arguments, we review the main theoretical
and experimental results that lead to the current physical picture of the creep regime.
In particular, we discuss recent works unveiling the collective nature of such ultra-slow
motion in terms of elementary activated events.
We show that these events control the mean velocity of the interface and cluster into
``creep avalanches'' statistically similar to the deterministic avalanches observed at
the depinning critical threshold.
The associated spatio-temporal patterns of activated events have been recently observed
in experiments with magnetic domain walls.
The emergent physical picture is expected to be relevant for a large family of disordered
systems presenting thermally activated  dynamics.
\end{abstract}

\maketitle

\tableofcontents

\section{Introduction}
\label{sec:introduction}

Our understanding of physics is largely based on idealized problems, the famous `spherical cows'.
Yet, the beauty of nature makes use of a much vast complexity.
It is well known nowadays that the presence of impurities and defects messing up with those rounded mammals 
leads to new emerging physical behavior, not observed in the idealized disorder-free problems. 
For example, the equilibration time of glasses becomes so large that it results to be experimentally
inaccessible. Such systems avoid crystallization and basically live
forever out-of-equilibrium~\cite{Barrat2003,Berthier2011}.
Dirty metals display localization and metal insulator transitions, unseen in perfect crystals~\cite{anderson1958absence,evers2008anderson}.
Systems of a broadly diverse nature show intermittent dynamics induced by
the presence of disorder~\cite{sethna2001crackling}.
Strained amorphous materials~\cite{baret2002extremal,lin2014scaling,nicolas2018deformation},
fracture fronts~\cite{bonamy2011failure,schmittbuhl1995interfacial,bonamy2008crackling},
magnetic~\cite{FerreCR2013,zapperi1998dynamics}
and ferroelectric domain walls~\cite{paruch2013nanoscale,kleemann2007universal},
liquid contacts lines~\cite{moulinet2004width,le2009height}, they all share a common phenomenology:
when the applied drive is just enough to induce motion, most of the system remains pinned
but large regions move collectively at high velocity.
These reorganizations are called avalanches.
Their location is typically unpredictable and their size distribution display a scale free statistics.
Given the ubiquity of this stick-slip behavior, the study avalanches has occupied a central scene
in non-equilibrium statistical physics, as can be seen in the large literature of sandpile models~\cite{dhar1999abelian},
directed percolation and cellular automata~\cite{henkel2008non}.

The depinning of an elastic interface moving in a disordered medium~\cite{narayan1993,nattermann1992,Fisher1998,MullerPRB2001,AgoritsasPhysB2012,FerreroCR2013}
is one of the paradigmatic examples where avalanches are well understood, thanks to the analogy with
standard equilibrium critical phenomena \cite{Fisher1998,Kardar1998}.
When the interface is driven at the force $f$ two phases are generically observed:
for $f<f_c$ the interface is pinned at zero temperature and motion is observed only during a transient time,
for $f>f_c$ the line moves with a finite steady velocity.
At $f_c$ the system displays a dynamical phase transition and the diverging size of avalanches
is the outcome of the presence of critical correlations.
Below and above $f_c$ the avalanches display a finite cut-off, that diverges approaching $f_c$.
We presently know the statistics of avalanches sizes~\cite{rosso2009avalanche}
and durations~\cite{kolton2019distribution} and their characteristic
shape~\cite{papanikolaou2011universality,laurson2013evolution}.
An important observation is that subsequent depinning avalanches are uncorrelated in space
and time at variance with the avalanche behavior observed in many systems where a `main-shock'
is at the origin of a cascade of `after-shocks'.
The so-called Omori law and productivity law, central in the geophysics of earthquakes~\cite{Scholz_2002},
are not present at the depinning transition~\footnote{Although depinning-inspired models have been adapted
to produce aftershocks by adding terms of slow relaxation or memory~\cite{JaglaJGR2010,JaglaPRL2014}}.
Namely all the experimental observations of depinning avalanches
temporally correlated were shown to be related to a finite detection threshold,
created by the limited sensitivity of the measurement apparatus~\cite{janicevic2016interevent}.

Nonetheless, genuine aftershocks could be
experimentally observed far from the depinning transition, in the so-called {\it creep} regime.
This regime, which describes the motion of magnetic domain walls at finite (e.g. room) temperature
and low applied fields, corresponds to an interface pulled by a small force
($f\ll f_c$) at finite temperature~\cite{KoltonPRL2006,AgoritsasPhysB2012,FerreroCR2013}.
The collective dynamics observed in this case is qualitatively different from the
one at the critical threshold.
In both regimes the dynamics is collective and involves large scale reorganizations.
But from the more recent results  creep ``avalanches" display complex
spatio-temporal patterns similar to the ones of observed in earthquakes.

In this paper we review the main arguments and results of the last thirty years about creep
with particular attention to the recent progress.
The paper is organized as follows.
In Sect.~\ref{sec:Tzerophasediagram} we introduce the model, present the dynamical regimes at zero temperature and
discuss the different universality classes.
In Sect.~\ref{sec:velocityatfiniteT} we provide the scaling arguments leading to
the creep law, namely the behavior of the steady velocity as a function of the applied force
at finite temperature. 
The numerical methods are discussed in Sect.~\ref{sec:numericalmethods}.
The more recent results valid in the limit of vanishing temperature are presented in Sect.~\ref{sec:numericalcreep}.
In Sect.~\ref{sec:experiments} we review the creep experiments on domain wall dynamics.
Conclusions and perspectives are given in Sect.~\ref{sec_Concl}.

\section{Dynamical phase diagram at zero temperature}
\label{sec:Tzerophasediagram}

\iffigures
\begin{figure}[!tb]
  \includegraphics[width=0.9\textwidth]{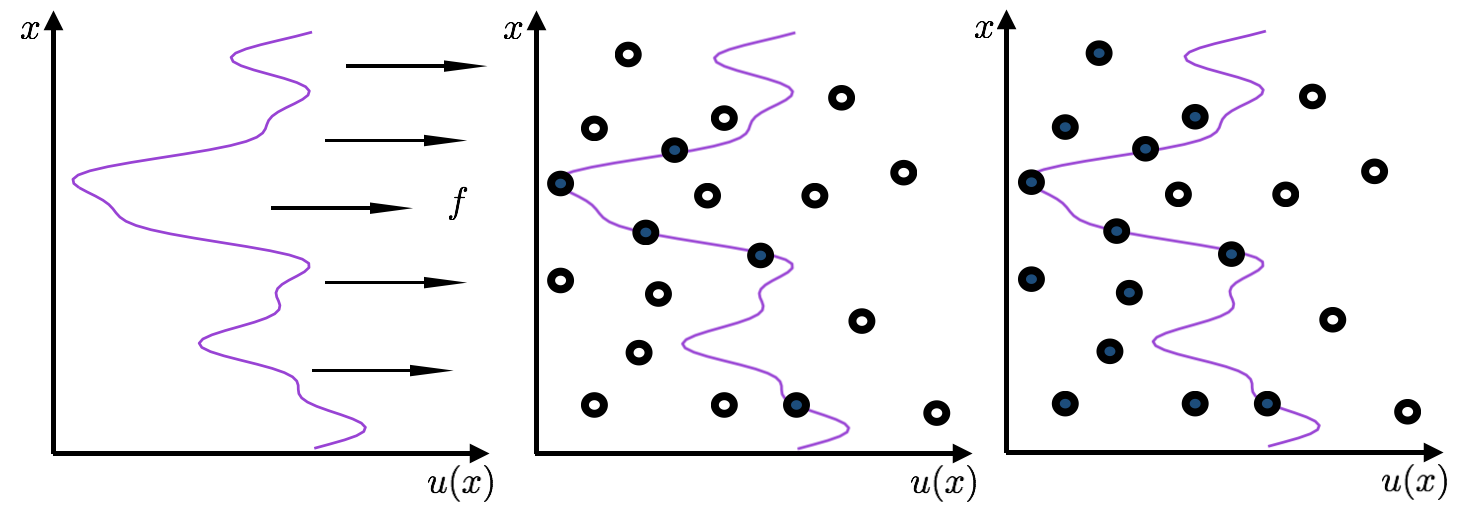}
  \caption{
  \textit{Left:} Sketch of the interface pulled by an external force $f$.
  The dark circles are the impurities that contribute to the pinning energy of the
  interface. In the random bond case  \textit{(center)} only neighboring impurities
  contribute while in the random field case  \textit{(right)} all
  the impurities on the left side of the interface contribute.}
  \label{fig:sketch_interface}
\end{figure}
\fi

We consider a $d$-dimensional interface in a ${d+1}$ disordered medium.
For simplicity we assume  that the local displacement at any time $t$
is described by a single valued function $u(x,t)$
(see \textbf{Figure \ref{fig:sketch_interface} left})
and that the dynamics is overdamped.
At zero temperature the equation of motion of the elastic manifold writes:
\begin{equation}
  \label{eq:qEWeom}
  \gamma \partial_t u(x,t)= c\nabla^2 u(x,t) + f + F_{\rm p}(x,u)
\end{equation}
where $c\nabla^2 u(x,t)$ describes the elastic force due to the surface tension,
$f$ is the external pulling force and $\gamma$ the microscopic friction.
The fluctuations induced by impurities are encoded in the quenched stochastic term
$F_{\rm p}=-\partial_u V_{\rm p}(x,u)$, where the energy potential $V_{\rm p}(x,u)$
describes the coupling between the manifold and the impurities.

For simplicity we assume the absence
of correlations along the $x$ direction \footnote{See Ref.\cite{Fedorenko2006} for a discussion of the correlated disorder case.}, while
the correlations of $V_{\rm p}(x,u)$ along the $u$
direction usually belong to one
of two universality classes:
(i) In the Random Bond class (RB) the impurities affect in a symetric way the the phases on each side of the interface. They thus simply locally attract or repel
the interface (see \textbf{Figure \ref{fig:sketch_interface} center}).
In this case the pinning potential and the pinning force are both
short-ranged correlated.
(ii) The Random Field class (RF) describes a disorder coupling in a different way in the two phases around the interface. Thus the 
pinning energies are affected by the
impurities inside the entire region delimited by the interface (see
\textbf{Figure \ref{fig:sketch_interface} right}).
Then $F_{\rm p}$ displays short range correlations while the pinning potential
$V_{\rm p}(x,u)$ displays long-range correlations
$\overline{[V_p(x,u)-V_p(x',u')]^2} \propto \delta(x-x')|u-u'|$. Here, the
overline denotes average over disorder realizations.

Equation Eq.\ref{eq:qEWeom},
so called quenched Edwards-Wilkinson equation,
is a coarse-grained minimal model governing the dynamics of the interface, at zero temperature for the moment,
at large scales~\cite{Fisher1998,Kardar1998,FerreroCR2013}.
It is  a non-linear equation in $u$ that
has been extensively studied by numerical simulation~\cite{FerreroPRE2013},
functional renormalization group techniques (FRG)~\cite{narayan1992,nattermann1992,ledoussal2002}
and exact mean-field solutions~\cite{fisher1985,alessandro1990,LeDoussalPhysC1995}.
For the case of a contact line of a liquid meniscus \cite{Joanny1984} as well as the crack
front of a brittle material \cite{Gao1989} the local elastic force is replaced by a long range one:
\begin{equation}
  \label{eq:elastic_force}
  c \nabla^2 u \;\rightarrow \;c \int \frac{(u(x',t)-u(x,t))}{|x'-x|^{\alpha+d}} {\rm d}^d x'
\end{equation}
with $\alpha=1$ and $d=1$.
The qualitative phenomenology of this generalized long range model is similar
to the quenched Edwards-Wilkinson, but the universal properties
(as critical exponents and scaling functions) are different.
However, for $\alpha\geq 2$ one recovers the short-range
universality class~\cite{KoltonJagla2018}.

The solution of this class of equations shows a behavior
reminiscent of second order phase transitions
with the velocity playing the role of the order parameter and the force
acting as the control parameter.
In particular, below a critical {\it depinning} threshold $f_c$ the steady velocity is zero,
and it acquires a finite value above only above that threshold.
The velocity vanishes continuously at the critical force as $v \simeq (f-f_c)^\beta$.
At the depinning the interface appears rough with a width
\begin{equation}\label{eq_roughness}
w^2(L) =  \frac{1}{L} \int_0^L u^2(x) {\rm d} x -\left(  \frac{1}{L} \int_0^L u(x) {\rm d} x \right)^2
\end{equation}
that grows as $L^{2\zetadep}$, 
with $L$ being the size of the system and $\zeta$ the roughness exponent.
Both $\beta$ and $\zetadep$ are universal depinning exponents
depending on the dimension $d$ of the interface and on the range $\alpha$ of
the elastic force; but interestingly, not on the disorder type~\cite{narayan1993,chauve2000}.
Slightly above $f_c$ the dynamics of a point of the interface is highly intermittent:
for long times the point is stuck with a vanishing velocity (much smaller than the average value $v$)
and suddenly starts to move with a high velocity. 
In equilibrium second order phase transition the universality arises from the
existence of a correlation length that diverges approaching the critical threshold.
For depinning the system is out-of-equilibrium but the presence
of large spatial correlations is manifested by the collective nature of this
intermittent dynamics: at a given time, while many pieces of the interface are at rest,
large and spatially connected portions move fast and coherently.

\iffigures
\begin{figure}[!tb]
  \includegraphics[width=0.9\textwidth]{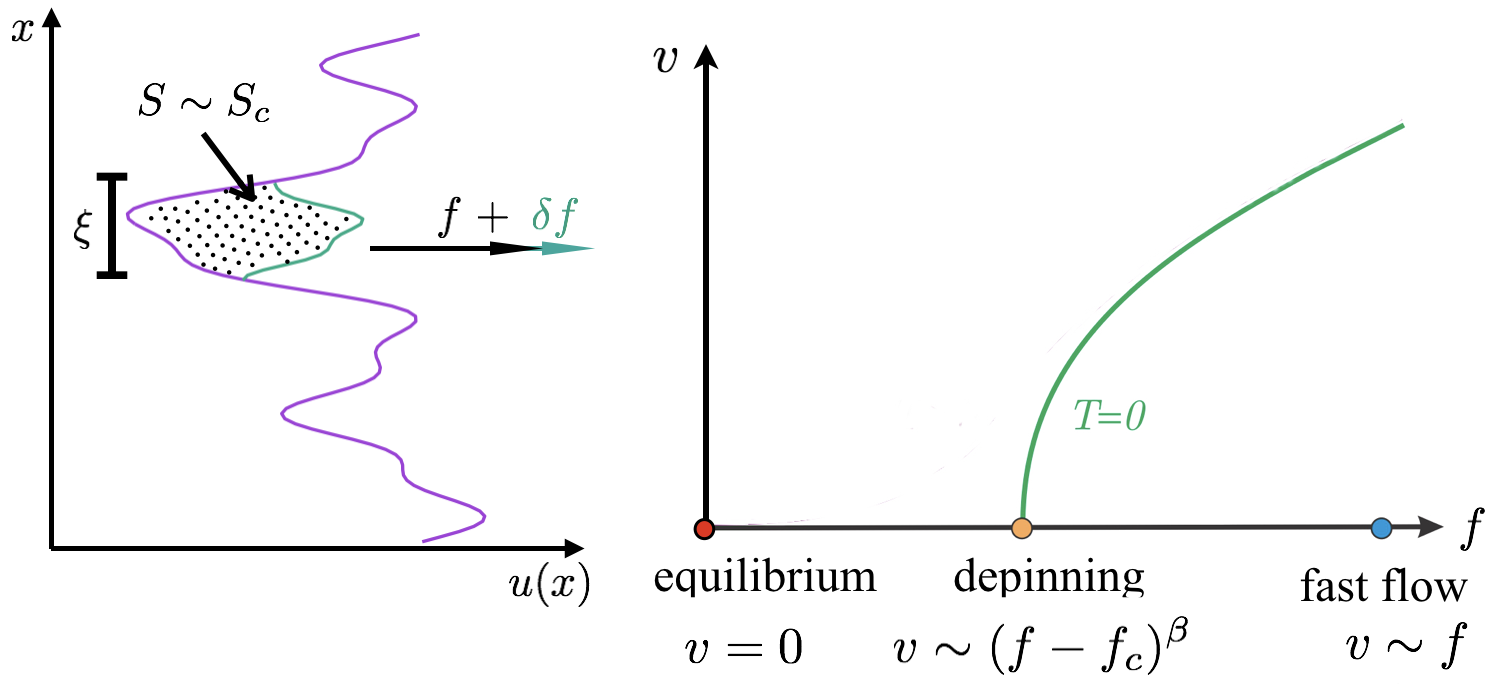}
  \caption{
  \textit{Left:} Sketch of an avalanche below $f_c$: the applied force $f$ is
  increased infinitesimally and a finite portion of the interface is destabilized.
  The size $S$ of the avalanche corresponds to the spanned area.
  \textit{Right:} Dynamical phase diagram at zero temperature.
  At $f=f_c$ the velocity and the shape of the interface have a universal
  scaling behavior, the dynamics is characterized by large and scale free avalanches.
  At $f=0$ the interface is in the ground state with a different roughness exponent
  which depends on the correlation of the disorder (RB or RF).
  At very large force the interface flows with a velocity that grows linearly with
  the force and the quenched disorder acts as a thermal noise.
  } \label{fig:phase_diagram}
\end{figure}
\fi

The presence of large correlations can be detected using a quasistatic protocol
below (but close to) $f_c$.
This is shown in \textbf{Figure \ref{fig:phase_diagram} left}
where an interface is at rest at a force $f$.
Upon increasing infinitesimally the force $f \to f + \delta f$, an avalanche takes place:
a large portion of the interface advances a finite amount while elsewhere
only readjusts infinitesimally ($\propto \delta f$).
The avalanches locations cannot be predicted 
and their sizes (the areas spanned between two consecutive metastable states) present
scale free statistics
\begin{equation}
P(S)=S^{-\taudep} g(S/S_c) \ .
\end{equation}
The Gutenberg-Richter exponent $\tau$ is universal as are $\beta$ and $\zetadep$,
$g(x)$ is a function that decays fast for $x\geq 1$ and is constant for $x<1$.
The characteristic size of the maximal avalanche increases
when $f\to f_c^-$.
In practice, $S_c$ is the clear manifestation of the divergent correlation
length $\xi \simeq |f-f_c|^{-\nudep}$ and one expects
$S_c \simeq \xi^{d+\zetadep} \simeq  |f-f_c|^{-\nudep(d+\zetadep)}$.
Many works have been devoted to describe the dynamics inside an avalanche~\cite{LeDoussalPRE2013,JaglaPRL2014,
janicevic2016interevent,kolton2019distribution,priol2019universal}:
typically the instability starts well localized at a given point and speads
in space over a distance $x(t) \simeq t^{1/z}$ up to a time $t_c \simeq \xi^z$.
For the qEW equation ~\ref{eq:qEWeom} it has been proven that there are only
two independent exponents, e.g. $\zetadep$ and $z$, and the other can be computed
by non trivial scaling relations (see \textbf{Table \ref{tab:dep}}).
Note that these relations are valid in low dimensions, because for $d\geq2\alpha$ the
value of the exponents saturates at their mean field value.

\begin{table}[t]
  \caption{
  Depinning exponents are known numerically with good precision
  and saturate to their mean field values for $d\geq 2\alpha$.
  At the depinning RB and RF disorder are in the same universality class.
  The numerical values of the roughness exponents $\zetadep$ are taken from
  \cite{Rosso2002} for $\alpha=1$ and from \cite{RossoPRE2003} for $\alpha=2$.
  Those of the dynamical exponent $z$ are taken from \cite{ramanathan1998}
  for $\alpha=1$ $d=1$, from \cite{leschhorn1993} for $\alpha=2$ $d=2$ and
  from \cite{FerreroPRE2013} for $\alpha=1$ $d=1$.\\
  }\label{tab:dep}
  \centering
  \begin{tabular}{cccccc}
  \hline
  \hline
  \textbf{Depinning}  &  \textbf{Observable} &  $d=1$   &  $d=1$   & $d=2$  & \textit{Mean Field}  \\
  \textbf{exponent}  & & $\alpha=2$ & $\alpha=1$ & $\alpha=2$  &  $d\geq 2\alpha$  \\
  \hline
  $z$ & $ t(L) \sim L^z$ & 1.43 & 0.77 & 1.56  & $\alpha$ \\
  $\zetadep$ & $u(x) \sim x^{\zetadep}$ & 1.25 & 0.39 & 0.75 & 0 \\
  $\taudep$ & $P(S) \sim S^{-\taudep}$ & \multicolumn{3}{c}{$\taudep=2-\alpha/(d+\zetadep)$} & 3/2 \\
  $\nudep$ & $\xi \sim |f-f_c|^{-\nudep}$ &   \multicolumn{3}{c}{$\nudep=1/(\alpha-\zetadep)$}  & $\alpha^{-1}$ \\
  $\beta$ & $v \sim |f-f_c|^{\beta}$ & \multicolumn{3}{c}{$\beta=\nudep(z-\zetadep)$}  & 1 \\
  \hline
  \hline
  \end{tabular}
\end{table}

The physics is very different in the limits of very small or very high forces.
At $f=0$ the interface is at equilibrium in the ground state,
its roughness is characterized by a very different (smaller) roughness exponent
and the nature of the disorder matters: RF interfaces are rougher than  RB.
The ground state energy is an extensive quantity (grows as $L^d$)  but its sample to
sample fluctuations scale as $L^{\theta}$.
The energy exponent $\theta$ obeys the scaling relation $\theta=2\zetaeq+d-\alpha$
(see \textbf{Table \ref{tab:eq}}).
This relation is a consequence of the statistical tilt symmetry of the model
which assures that the elastic constant $c$ is not renormalized.
On the other hand, assuming that in equilibrium elastic and disorder energy scale in the same way, one has
from $E_{\rm el}[u] = \frac{c}{2} \int
\frac{(u(x',t)-u(x,t))^2}{|x'-x|^{\alpha+d}} {\rm d}^d x {\rm d}^d x'$
the relation $E_{\rm eq} \propto L^{2\zetaeq}L^{-(\alpha+d)}L^2 \sim L^{2\zetaeq+d-\alpha}$.
Note that for $\alpha>d/2$, the interface is flat ($\zetaeq=0$) and the energy exponent saturates
to the central limit value $\theta=d/2$.

At $f\to\infty$ the quenched pinning reduces to an annealed stochastic noise
because in the comoving frame one has $F_{\rm p}(x,u)=F_{\rm p}(x,\delta u+ v t) \sim F_{\rm p}(x,vt)$.
For short-range correlated pinning force, the strength of the disorder plays the role of
and effective temperature $T_{\rm eff}$.
In this so-called fast-flow regime the motion is not intermittent, and
one recovers the standard Edwards Wilkinson dynamics with the generalized
fractional laplacian of Eq.\ref{eq:elastic_force} \cite{Zoia2007}.
In particular the dynamical exponent is $z=\alpha$ and the roughness
exponent is $\zeta_{\rm flow}=(\alpha-d)/2$ for $d\leq\alpha$.
For larger dimension, the Edwards Wilkinson interface is flat.

For intermediate forces the physics is not fully governed
by any of the three characteristic points described above ($f=f_c$, $f=0$ and $f\to\infty$).
Therefore, one could wonder if a completely new scaling description should be introduced.
It turns out that it is not the case, at least for $f>f_c$.
The physics of the interface can be described by a crossover between short length scales,
governed by the critical behaviour at $f=f_c$, and large length scales, governed by the
fixed point of $f=\infty$.
Below the depinning threshold, $f<f_c$, no steady-state can be defined at zero temperature
rather than the complete arrest of the interface.
The presence of a finite temperature, discussed in the next section, allows to investigate a
non-trivial stationary dynamical regime (the creep) with finite velocity at forces in between
the equilibrium and the depinning fixed point, and to analyze how this two fixed points
affect the dynamics at different scales.

\begin{table}[t]
  \caption{Equilibrium exponents for elastic manifold with random bond disorder (RB).
  For $\alpha=2$ the results in $d=1$ are exact.
  In $d=2$ we used the numerical results from~\cite{middleton1995}
  obtained using a maximal flow algorithm.
  For $\alpha=1$ the results are known from FRG calculations, for RF disorder
  one expects $\zetaeq=\theta=1/3$. Note that $\theta$ and $\zetaeq$ are not independent, but obey to
  the following scaling relation $\theta=2\zetaeq+d-\alpha$.\\
  }\label{tab:eq}
  \centering
  \begin{tabular}{cccccc}
    \hline
    \hline
   \textbf{Equilibrium}  &  \textbf{Observable} &  $d=1$   &  $d=1$   & $d=2$ & \textit{Mean Field}  \\
   \textbf{exponent}  & & $\alpha=2$ & $\alpha=1$ & $\alpha=2$  & $d\geq 2\alpha$ \\
    \hline
    $\theta$ & $ E(L) \sim L^\theta$ & 1/3 & $\simeq$ 0.2 & $\simeq$ 0.84 & $d/2$ \\
   $\zetaeq$ & $u(x) \sim x^{\zetaeq}$ & 2/3 & $\simeq$ 0.2 & $\simeq$ 0.41 & 0 \\
   $\taueq$ & $P(S) \sim S^{-\taueq}$ & \multicolumn{3}{c}{$\taueq=2-\alpha/(d+\zetaeq)$} & 3/2  \\
   $\nueq$ & $\xi \sim f^{-\nueq}$ &   \multicolumn{3}{c}{$\nueq=1/(\alpha-\zetaeq)$}  & $\alpha^{-1}$\\
    \hline
    \hline
  \end{tabular}
\end{table}

\subsection{The case of the quenched Kardar-Parisi-Zhang (KPZ) depinning}
\label{sec:thecaseofqKPZ}

\begin{table}[b]
  \caption{Exponents of the qKPZ depinning universality class.
  The numerical values of the roughness exponent $\zetadep$ are taken from \cite{RossoPRE2003}.
  For $d=1$ the exponents $z$ and $\nudep$ are taken from \cite{Tang1995},
  while for $d=2$ from \cite{Buldyrev1993}.
  The existence of an upper critical dimension is under debate.\\
  }\label{tab:qKPZ}
  \centering
  \begin{tabular}{cccc}
  \hline
  \hline
  \textbf{qKPZ}  &  \textbf{Observable} &  $d=1$    & $d=2$   \\
  \textbf{exponent}  & & $\quad\alpha=2\quad$ & $\alpha=2$  \\
  \hline
  $z$ & $ t(L) \sim L^z$ & 1 & 1.1 \\
  $\zetadep$ & $u(x) \sim x^{\zetadep}$ & 0.63 & 0.45 \\
  $\nudep$ & $\xi \sim |f-f_c|^{-\nudep}$ & 1.733 & 1.05 \\
  $\taudep$ & $P(S) \sim S^{-\taudep}$ & \multicolumn{2}{c}{$\taudep=2-(\zetadep+1/\nudep)/(d+\zetadep)$}  \\
  $\beta$ & $v \sim |f-f_c|^{\beta}$ & \multicolumn{2}{c}{$\beta=\nudep (z-\zetadep)$}  \\
  \hline
  \hline
  \end{tabular}
\end{table}

The quenched Edwards Wilkinson equation and its generalization
to long range elasticity are well studied and understood.
In all these models the non-stochastic part of the equation is
linear in the displacement $u$ and one can derive the scaling
relation of table \ref{tab:dep}.
However, in presence of anisotropies in the disorder~\cite{Tang1995}
or in the elastic interaction~\cite{RossoPRL2001}, a non-linearity
becomes relevant for short range elasticity.
In this case the equation of motion of the interface writes:
\begin{equation}\label{eq:qKPZeom}
\gamma \partial_t u(x,t)= c\nabla^2 u(x,t) + \lambda (\nabla u(x,t))^2+ f + F_{\rm p}(x,u) \ .
\end{equation}

The inclusion of this non-linear term affects the physical behavior
as $f\to\infty$ leading to the standard Kardar-Parisi-Zhang (KPZ) ~\cite{kardar1986dynamic}
dynamics rather than the Edwards Wilkinson.
At depinning, if $\lambda f \geq 0$ the motion remains intermittent with large
avalanches but with different exponents \cite{Tang1992,Buldyrev1993}
characterized by new scaling relations, as shown in \textbf{Table \ref{tab:qKPZ}}.
When $\lambda f <0$ the interface develops a sawtooth shape with an effective exponent
$\zetadep=1$~\cite{Jeong1996}.
This regime has been recently observed in~\cite{Atis2015}.

\section{Velocity at finite temperature}
\label{sec:velocityatfiniteT}

At finite temperature the interface has a finite steady velocity $v$,
even below $f_c$.
The energy of the interface can be written as the sum of three
contributions:
\begin{equation} \label{eq:E}
  E[u] =  \int_0^L {\rm d}^d x \  \left[  \frac{c}{2}(\nabla u(x) )^2 +  V_{\rm p}(x,u(x)) - f   u(x) \right] \ ,
\end{equation}
the first term on the {\tt RHS} being the elastic energy of the interface,
the second, the pinning potential, and the third, the energy associated
to the driving force $f$.
We note that the equation of motion (\ref{eq:qEWeom}) is obtained from
$\gamma \partial_t u(x,t) = - \delta E[u] / \delta u(x,t)$.
At finite temperature one can write the associated Langevin equation:
\begin{equation}
\label{eq:dynamics_finiteT}
\gamma \partial_t u(x,t)= c\nabla^2 u(x,t) + f + F_{\rm p}(x,u) + \eta(x,t) \ ,
\end{equation}
with $\langle \eta(x,t)\eta(x',t') \rangle = 2 \gamma T \delta(t-t')\delta(x-x')$
where the average is over different realizations of the thermal noise, while the
disordered landscape remains fixed.

In presence of a finite drive, the energy Eq.~\ref{eq:E} has no lower bound as
it is tilted by the force and in average decreases linearly by increasing $u$.
Yet, the presence of pinning generates metastable states and barriers up to $f_c$.
The activated motion at finite temperature allows to overcome these barriers
yielding a finite steady-state velocity.

\iffigures
\begin{figure}[!t]
  \includegraphics[width=0.9\textwidth]{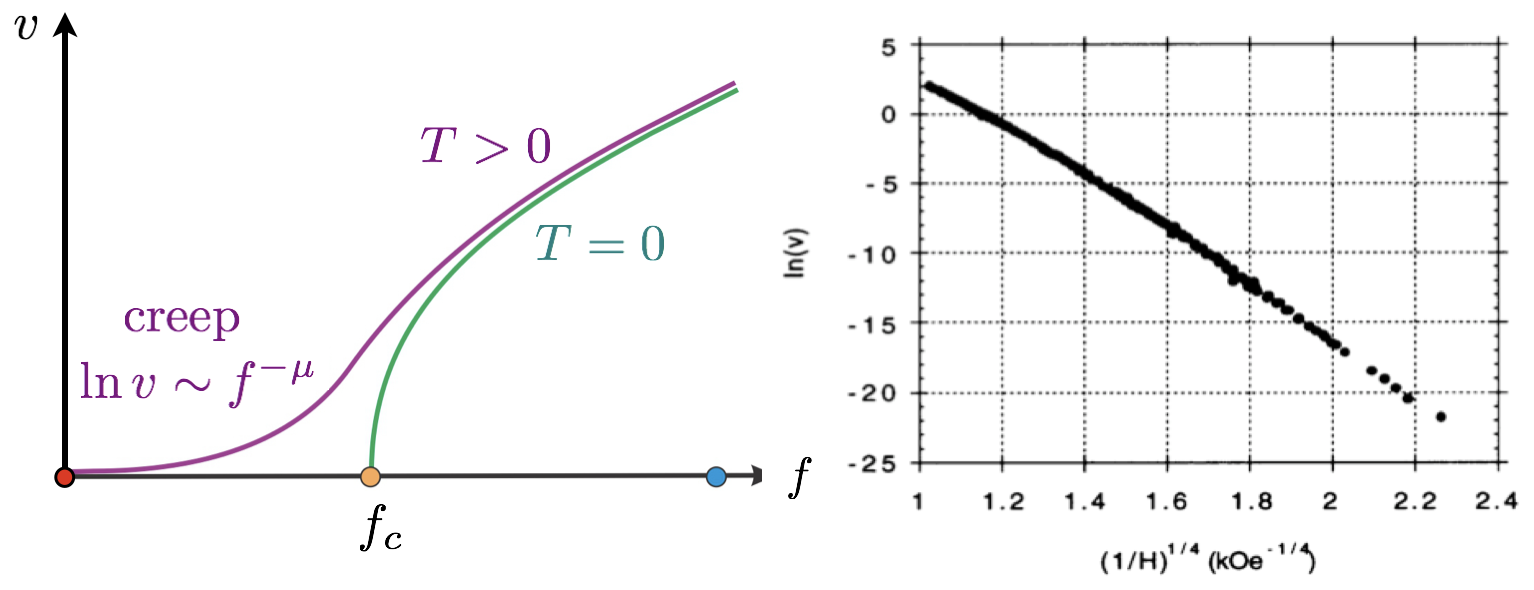}
  \caption{
  \textit{Left:} Velocity force characteristics at finite temperature.
  When $f$ is very small compared to $f_c$ and at very small temperature,
  one observes the creep law $\ln v \sim f^{-\mu}$. Adapted from \cite{FerreroCR2013}.
  \textit{Right:} First experimental verification of a creep law consistent with $\mu=1/4$
  in 2d ultra-thin Pt/Co/Pt film at room temperature, taken from \cite{LemerlePRL1998}.
  }
  \label{fig:creep}
\end{figure}
\fi

The velocity force characteristics is represented in \textbf{Figure \ref{fig:creep} left}.
At very small force and finite temperature a creep regime is observed, where the velocity
displays a stretched exponential behavior:
\begin{equation}\label{eq:creep}
  \displaystyle v(f,T) = v_0 e^{- \left(\frac{f_T}{f}\right)^\mu } \ ,
\end{equation}
with $v_0$ and $f_T$ depending on the temperature and the microscopic parameters,
while $\mu$ is a universal exponent.
This creep law was verified experimentally in ferromagnetic ultrathin films with $\mu\simeq 1/4$
first by Lemerle \textit{et al.} \cite{LemerlePRL1998} (see  \textbf{Figure \ref{fig:creep} right}).
Rather strikingly, this law can span several decades of velocity
(from almost walking speed to nails growth speed) by just varying one decade of the
externally applied magnetic field at ambient temperature.
The creep law was subsequently found by many other experiments\cite{KimNature2009,JeudyPRL2016}
(see Section \ref{sec:experiments} for a brief review), confirming the universality and robustness
of several creep properties.
Such universality naturally calls for minimal statistical-physics models on which we will focus.

Eq. \ref{eq:creep} has been predicted in \cite{Ioffe1987,Nattermann1987,vinokur1996}
and derived within the functional renormalization group technique in \cite{chauve2000}.
The stretched exponential behavior originates from the collective nature
of the low temperature dynamics of these extended objects.
For a point-like system embedded in a short-range disorder potential
the response to a small force will be linear in $f$.
The idea is to consider that the energy landscape is characterized
by valleys at distance $\Delta u$ separated by an energetic barrier
of typical size $E_{{\rm p}}$.
In presence of the tilt introduced by a finite force $f$, the energy
gap between two consecutive valleys becomes $\sim f \Delta u$
(see  \textbf{Figure \ref{fig:particle}}).
According to the Arrhenius law, the time to jump from left to right will
be $e^{\beta ( E_{{\rm p}} - f \Delta u/2) } $, while the time
for doing it from right to left would be $e^{\beta( E_{{\rm p}} + f \Delta u/2)}$.
Therefore, the velocity can be computed as the thermally assisted
flux flow (TAFF \cite{Anderson1964}) across the barrier:
\begin{equation}\label{eq:TAFF}
\displaystyle v \propto  e^{- \beta ( E_{{\rm p}} - f \Delta u/2) } - e^{- \beta( E_{{\rm p}} + f \Delta u/2)} \simeq e^{- \beta E_{{\rm p}}}  \Delta u f \ .
\end{equation}
We conclude that, in presence of bounded barriers, the velocity will be linear
even if with an exponentially suppressed mobility.

\iffigures
\begin{figure}[t]
  \begin{minipage}{.5\textwidth}
  \includegraphics[width=0.95\textwidth]{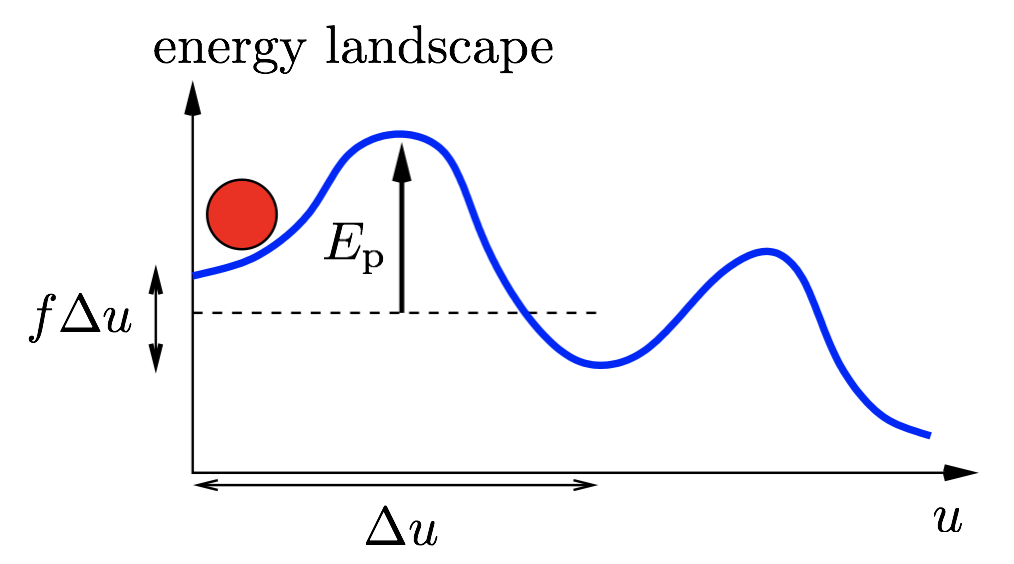}
  \end{minipage}
  \begin{minipage}{.5\textwidth}
  \includegraphics[width=0.95\textwidth]{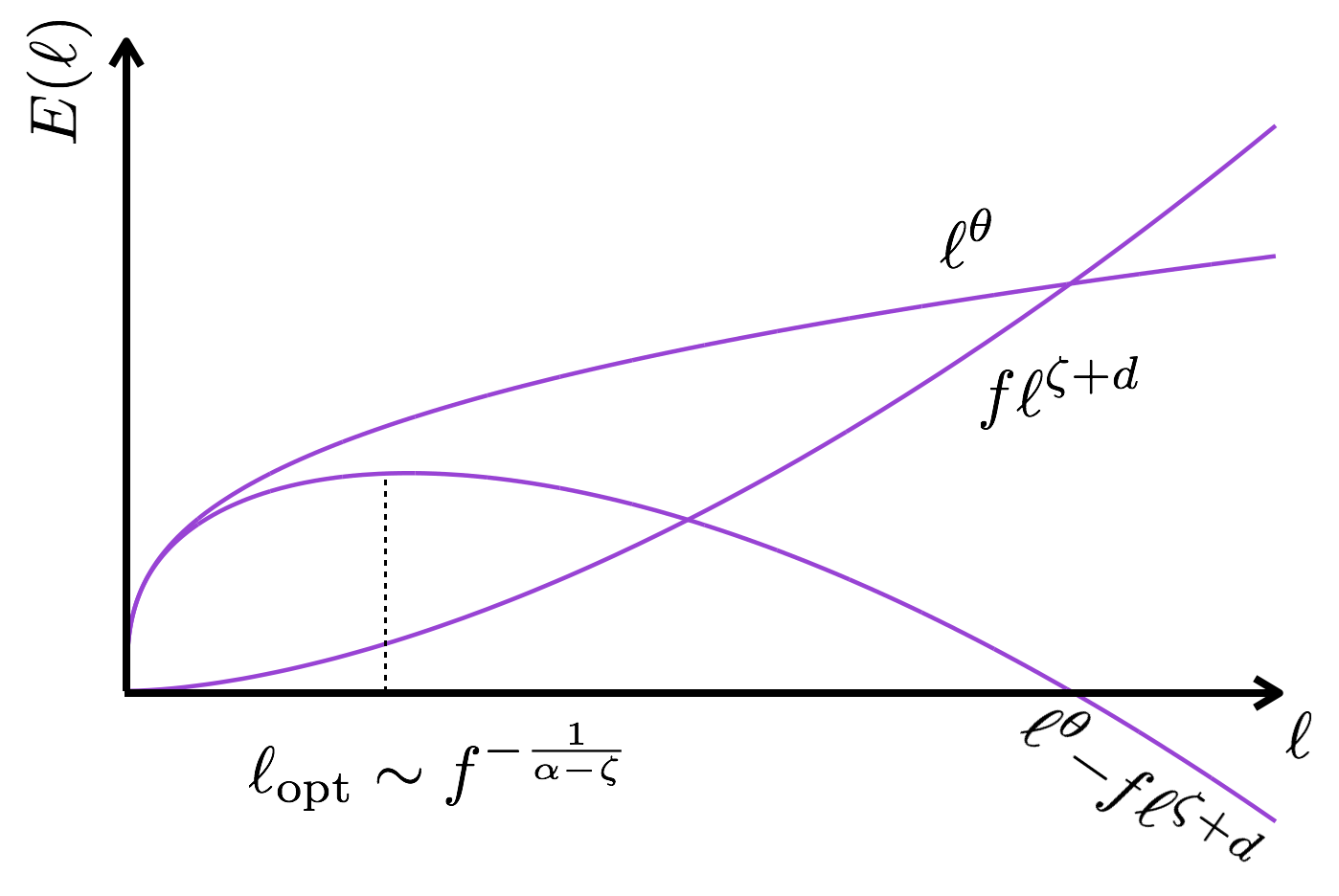}
  \end{minipage}
  \caption{
  \textit{Left:} Thermally assisted flux flow.
  The activated velocity of a single degree of freedom in a short range disordered potential
  is linear in the force and exponentially suppressed by the size of the typical barrier $E_{{\rm p}}$.
  \textit{Right:} Creep behavior.
  The energetic barrier encountered by an interface diverges when the applied force vanishes.
  Indeed in order to find a new metastable state characterized by smaller energy a large
  portion of the interface has to reorganize.
  Scaling arguments predict that the linear size of such reorganization scales as
  $\ell_{{\rm opt}} \sim f^{- \frac{1}{\alpha-\zetaeq}}$.}
  \label{fig:particle}
\end{figure}
\fi

\iffigures
\begin{wrapfigure}{R}{0.33\textwidth}
  \includegraphics[width=\linewidth]{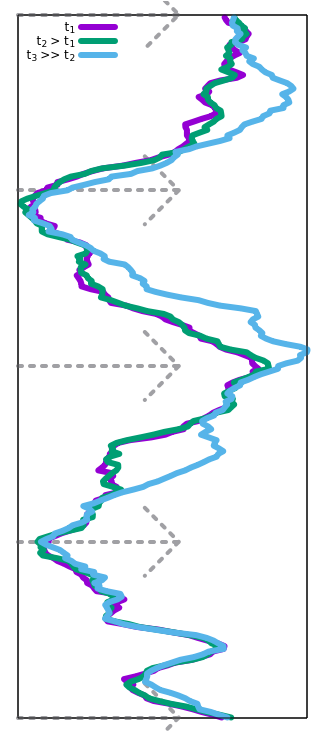}
  \caption{ \label{fig:futile}
  Configurations at different times obtained by direct integration of
  Eq.~\ref{eq:dynamics_finiteT}.
  At short times one observes incoherent oscillations and the configurations
  differ only at short length scales.
  At much larger times the line advances in the direction of the force
  with a coherent excitation that involves a large reorganization.}
\end{wrapfigure}
\fi

For an extended object the typical barrier grows when the external force vanishes and
their divergence is at the origin of the stretched exponential behavior in Eq. \ref{eq:creep}.
In  \textbf{Figure \ref{fig:futile}} we show different configurations obtained at different times
from the direct integration of Eq. \ref{eq:dynamics_finiteT}.
At short times one observes incoherent oscillations and the configurations differ
only at short length scales.
At much larger times the line advances in the direction of the force
with a coherent excitation that involves a large reorganization.
This collective motion leads the system to a local minimum characterized by
a  lower energy due to the presence of the force.
It is very unlikely that the interface will climb back to the previous configurations
characterized by a higher energy.
This new and deeper valley is the starting point of a new search in the forward direction.
At these time scales the dynamics of the line can be seen as a sequence
of metastable states
\begin{equation}
  \alpha_1 \to \alpha_2 \to \alpha_3 \to \dots
\end{equation}
characterized by decreasing energies
\begin{equation}
  E_{\alpha_1} > E_{\alpha_2} > E_{\alpha_3} > \dots
\end{equation}
At low temperature for a given $\alpha_1$, $\alpha_2$ is the metastable state with lower energy
 that can be reached crossing the minimal barrier.
It is possible to show that  for an interface of internal dimension $d$ embedded in a $d+1$ dimension
the pathway obtained with such a rule is the optimal one (and thus the one that dominates
the statistics of the dynamics) in the low temperature limit \cite{KoltonPRB2009}.

The first attempts to evaluate the barriers and the length scales associated
to this coarse grained dynamics have been done in \cite{Ioffe1987,Nattermann1987} and in \cite{chauve2000} via FRG.
The main assumption in their original derivation is that, during the dynamical evolution,
the energy barriers scale as the energy fluctuations of the ground state at $f=0$.
At equilibrium the fluctuations of the free energy are known to grow with the system size
with a characteristic exponent $\theta$ that depends on the equilibrium roughness exponent
via an exact scaling relation $\theta=2\zetaeq+d-\alpha$.
Numerical simulations in \cite{drossel1995} have shown that the barriers separating two equilibrium
metastable states, that differ on a portion $\ell$, grow as $\ell^\psi$ with an exponent consistent
with $\psi\simeq\theta$.
Using these ideas one can assume that the energy barriers due to the pinning centers
and in absence of tilt grow with the size of the reorganization
\begin{equation}\label{eq:Ep}
  E_{{\rm p}}(\ell) \sim \ell^{\theta} = \ell^{2\zetaeq+d-\alpha}
\end{equation}
If the motion is in the forward direction one has to subtract the energy induced by
the tilt
\begin{equation}\label{eq:Ef}
  E_{{\rm f}}(\ell) \sim f \, u(\ell) \, \ell^d = f \ell^{\zetaeq+d}
\end{equation}
In \textbf{Figure \ref{fig:particle} right} we show that the competition between
these two terms (Eqs.\ref{eq:Ep} and \ref{eq:Ef}) yields the characteristic length
scale $\Lopt$ of the optimal reorganization (and the optimal barrier $E_{{\rm p}}(\ell_{{\rm opt}})$)
allowing to reach a new metastable state with a lower energy:
\begin{equation}\label{eq:Lopt}
  \ell_{{\rm opt}} \sim f^{-\frac{1}{\alpha-\zetaeq}} \qquad E_{{\rm p}}(\ell_{{\rm opt}}) \sim f^{-\frac{\theta}{\alpha-\zetaeq}}.
\end{equation}
Using the scaling of $E_{{\rm p}}$ in Eq. \ref{eq:TAFF} one recovers the creep law, Eq. \ref{eq:creep}, and identifies
the creep exponent
\begin{equation}
  \mu = \frac{\theta}{\alpha-\zetaeq} = \frac{2\zetaeq+d-\alpha}{\alpha-\zetaeq}
\end{equation}
as an equilibrium exponent.
In particular in $d=1$, for RB disorder and short range elasticity one recovers
$\mu=1/4$ as in the experiment \cite{LemerlePRL1998}.

Although for the average velocity there is an excellent agreement between the simple
scaling arguments \cite{Ioffe1987,Nattermann1987} and the more sophisticated FRG
analysis \cite{chauve2000}, the FRG showed clearly that other lengthscales besides
$\ell_{{\rm opt}}$ (see Figure \ref{fig:particle} right) were necessary to describe
the motion, pointing to a rich dynamics in the creep regime.
In particular the FRG showed that the thermal nucleus led in the dynamics to
avalanches at a larger lengthscales than $\ell_{{\rm opt}}$ itself. 
In order to make a full analysis of the creep regime, a numerical investigation was
thus eminently suitable.
This is however a highly non-trivial task considering the exponentially large time
and length scales.
We discuss on how to undertake such a study in the next section.  

\section{Numerical methods}
\label{sec:numericalmethods}

The direct simulation of the Langevin equation \ref{eq:dynamics_finiteT}
has been performed in \cite{vinokur1996} and later in \cite{kolton2005creep}.
This approach confirms a non-linear behavior for the velocity-force
characteristics but fails in probing the specific scaling of the creep law.
In fact, at low temperature these methods can focus only on the microscopic
dynamics describing incoherent and futile oscillations around local minima
(see \textbf{Figure \ref{fig:futile}}).
The forward motion that allows to escape from these minima occurs at
very long time scales that are difficult to reach.
In practice one has to increase the temperature or the force bringing the
system beyond the validity of the creep scaling hypothesis.

A completely different strategy focus on the coarse grained dynamics at the
time scales of the coherent reorganizations that are able to lower the energy.
In practice one has to model the interface as a directed polymer
of $L$ monomers at integer positions $u(i)$, $i=1,\dots,L$ and with periodic
boundary conditions ($u(L+1)=u(1)$).
The energy of the polymer is given by:
\begin{equation} \label{eq:energy_discrete}
  E = \sum_i \left[ (u(i+1)- u(i))^2  - f u(i) + V(i,u(i)) \right].
\end{equation}
To reduce the configuration space it is useful to implement a
hard metric constraint such that
\begin{equation} \label{eq:hardconstraint}
  |u(i+1)-u(i)|\leq \kappa ,
\end{equation}
with $\kappa \sim \mathcal{O}(1)$ an integer.

To model RB disorder one can define $V_{RB}(i,u)=R_{i,u}$ with $R_{i,u}$ Gaussian
random numbers with zero mean and unit variance,
while for RF disorder
$V_{RF}(i,u)=\sum_{k=0}^{u} R_{i,k}$,
such that $\overline{[V_{RF}(i,j)-V_{RF}(i',j')]^2}=\delta_{i,i'} |j-j'|$.

At the coarse grained level the dynamics corresponds to a
sequence of polymer positions determined using a two step algorithm.
\begin{itemize}
\item {\it Thermal activation.}  Starting from any metastable state
one has to find the compact rearrangement that decreases the energy
by crossing the minimal barrier among all possible pathways. 
\item {\it Deterministic relaxation.} After the above activated move,
the polymer is not necessarily in a new metastable state and relaxes
deterministically with the non local Monte Carlo elementary moves
introduced in~\cite{RossoPRB2001}.
\end{itemize}
From the computational point of view the most difficult task is in the first step.
In principle, one fixes a maximal barrier and enumerates all possible pathways that
stay below the maximal allowed energy.
If one of them reaches a state with a lower energy the thermal activation step is over,
otherwise the maximal barrier is increased and the process is repeated.
This protocol is exact, it has been implemented in \cite{KoltonPRB2009},
but it has severe computation limitations at low forces as
the minimal barrier is expected to diverge for vanishing forces.
In order to explore the low force regime, a different strategy
has been adopted in \cite{FerreroPRL2017}.
Instead of looking to the pathway with the minimal barrier one selects the smallest
rearrangement that decreases the energy.
This is done by fixing a window $w$ and computing the optimal
path between two generic points $i,i+w$ of the polymer using
the Dijktra's algorithm adapted to find the minimal energy
polymer between two fixed points.
The minimal favorable rearrangement corresponds to the minimal window
for which the best path differs from the polymer configuration.
Using this strategy, it was possible not only to increase of a factor $30$
the system size, but, and more importantly, to decrease of a factor $100$
the external drive $f$, unveiling the genuine creep dynamics.

\section{Creep dynamics in the limit of vanishing temperature}
\label{sec:numericalcreep}

\iffigures
\begin{figure}[h]
  \includegraphics[width=0.85\textwidth]{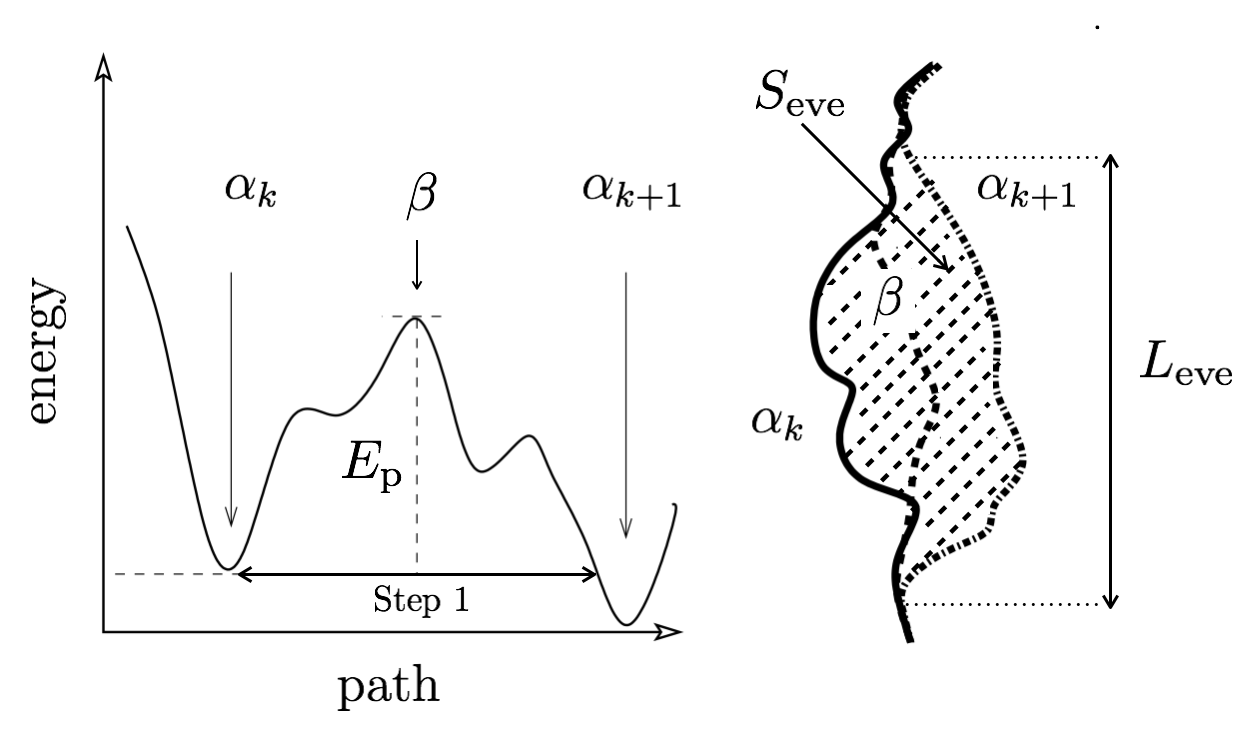}
  \caption{
  Sketch of the selected pathway starting from the metastable state $\alpha_k$.
  During `step one' of the algorithm one searches for a polymer configuration
  with an energy smaller than the one associated to $\alpha_k$ by crossing a minimal barrier $E_p$.
  During `step two' the polymer relaxes deterministically to a metastable configuration,
  no barriers are overcome at this stage.
  Adapted from \cite{KoltonPRB2009}.
  } \label{fig:path}
\end{figure}
\fi

Here we give a summary of the main results obtained using
the coarse grained dynamics introduced in \cite{KoltonPRB2009,FerreroPRL2017}.
The output of the algorithm is a sequence of metastable states $\alpha_k$
($k=1,\dots,n$), as shown in \textbf{Figure \ref{fig:path}}.
In \cite{KoltonPRB2009} the barrier $E_p$ is the minimal between all possible
pathways, while in \cite{FerreroPRL2017} the criterium of the minimal barrier
has been approximated with the criterium of the minimal rearrangement which
allows to reach much smaller forces and much larger sizes.
The area between two subsequent metastable states (see \textbf{Figure \ref{fig:path}})
defines the size of an activated event.
Below this size the dynamics is futile characterized by incoherent vibrations,
while once the new metastable state is reached the backward move is suppressed.

\subsection{Statistics of the events and clusters}
\label{sec:statisticsofevents}

From the scaling arguments of Section~\ref{sec:velocityatfiniteT} one expects that
the area of the activated events is of the order $\Lopt^{d+\zetaeq}$ with $\Lopt$
that grows when the force decreases (see Eq. \ref{eq:Lopt}).
However the distribution shown in \textbf{Figure  \ref{fig:events}}
displays a power law scaling analogous to the depinning one
\begin{equation} \label{eq:PSeve}
 P(\Seve) \sim S_{\rm eve}^{-\tau} g(\Seve/S_c) \ .
\end{equation}
When the force decreases the cutoff $S_c(f)$
grows and displays the scaling predicted in Section~\ref{sec:velocityatfiniteT}:
\begin{equation}
  S_c  \sim \Lopt^{d+\zetaeq} \sim f^{-\nueq (d+\zetaeq)} \ .
\end{equation}
Here $d=1$ and $\zetaeq$ depends on the nature of the
disorder: for RB $S_c(f) \sim f^{-5/4}$ while for RF $S_c(f) \sim f^{-2}$.

\iffigures
\begin{figure}[h]
  \begin{minipage}{.55\textwidth}
  \includegraphics[width=\textwidth]{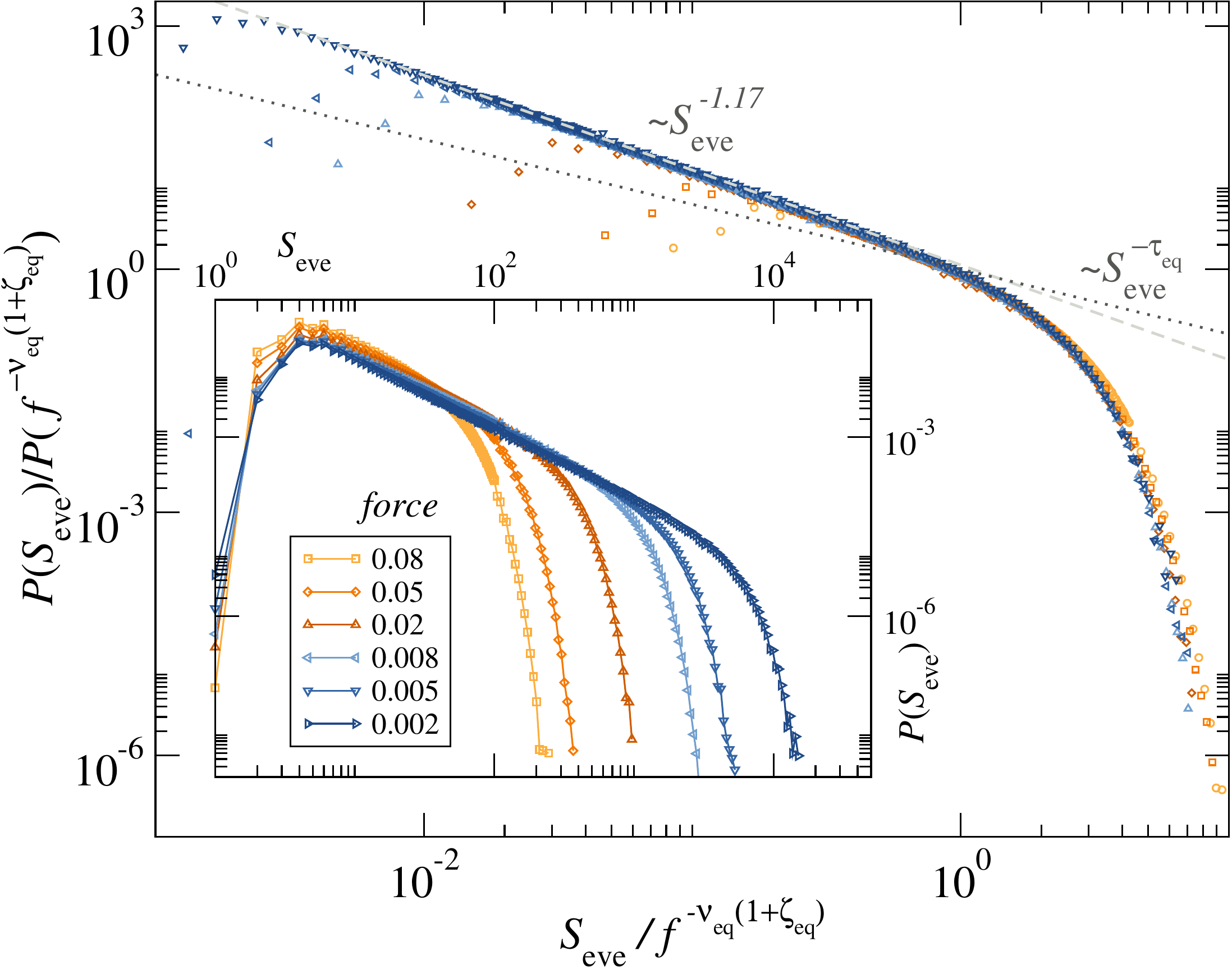}
  \end{minipage}
  \begin{minipage}{.45\textwidth}
  \includegraphics[width=\textwidth]{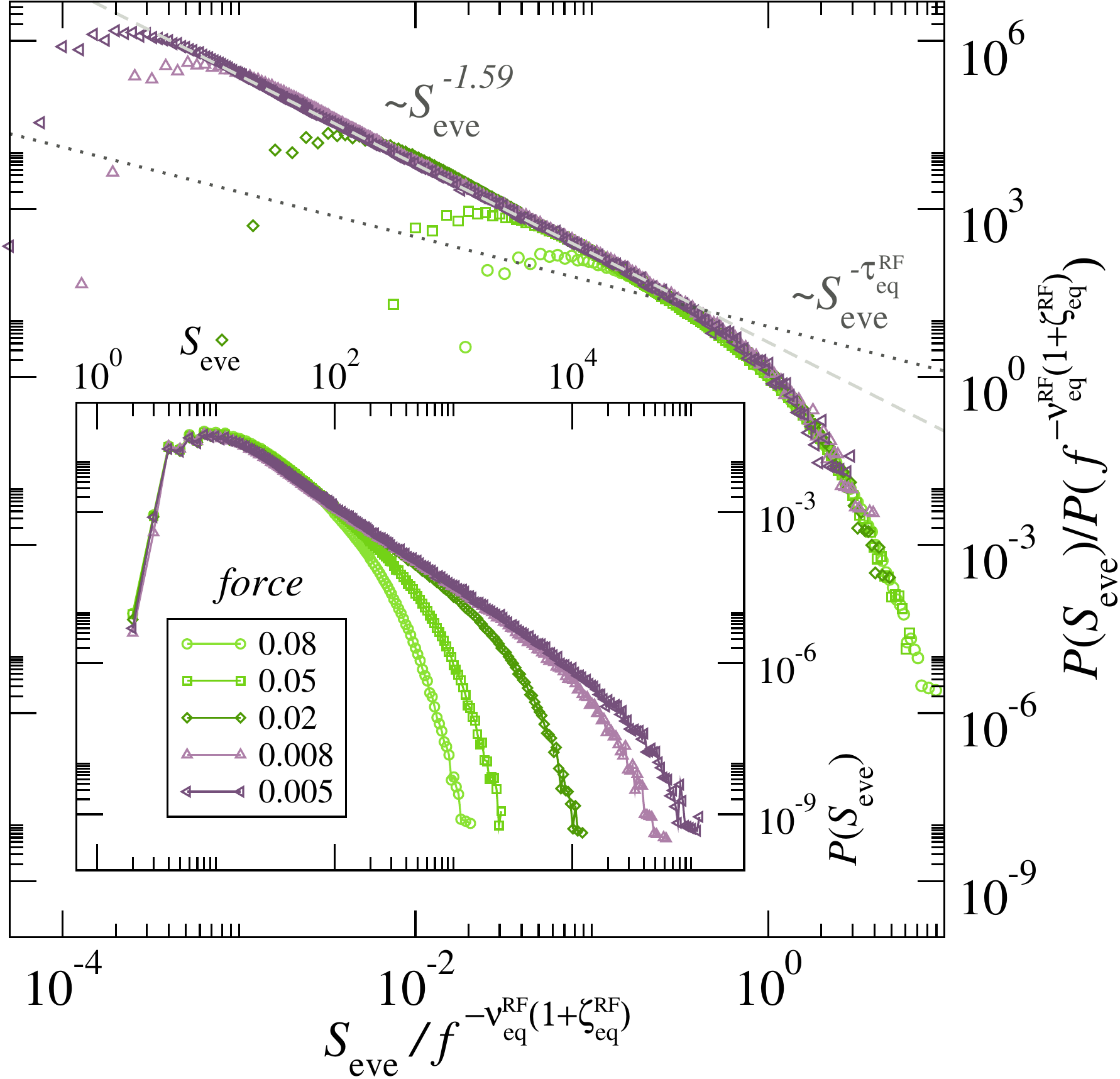}
  \end{minipage}
  \caption{
  Events size distributions $P(S_{\rm eve})$ for RB \textit{(left)} and RF \textit{(right)}
  at different forces.
  Main pannels show collapses by plotting $\Seve/S_c$ with $S_c(f)=f^{-\nueq(1+\zetaeq)}$.
  Insets show the unscaled distributions.
  Note that for RB disorder $S_c(f)=f^{-5/4}$ while for RF disorder $S_c(f)=f^{-2}$.
  The perfect collapse validates the expected creep scaling $\ell_{\rm opt} \sim f^{-\nueq}$,
  given $S_c\sim \ell_{\rm opt}^{(1+\zetaeq)}$.
  Adapted from \cite{FerreroPRL2017}.
  } \label{fig:events}
\end{figure}
\fi

Eq. \ref{eq:PSeve} implies that the typical activated events are much smaller
than the one predicted by scaling arguments.
However few very large events dominate the characteristic time scales of the
forward motion.
The behavior of the velocity in the creep formula is then determined by the
occurrence of such large reorganizations.
Indeed, the barriers associated to the largest elementary events are expected
to scale as $U_{\text opt}(f) \sim \Lopt^\theta \approx S_c(f)^{\theta/(d+\zetaeq)}$.
Then the mean velocity in the Arrhenius limit writes as $v \sim \exp[-U_{\text opt}/T]
\sim \exp[-(f_T/f)^\mu/T]$, with $\mu=\theta/(2-\zetaeq)$,
recovering the celebrated creep law of Eq. \ref{eq:creep}.
The main difference with the previous scaling  approaches \cite{Ioffe1987,Nattermann1987}
is that the creep law is not determined by the `typical' events
but by the largest ones instead.

To get further inside on the sequence of these events one notes that the
exponent $\tau$ of $P(S_{\rm eve})$ is larger than the one expected in equilibrium
(in particular in \textbf{Figure  \ref{fig:events}} for RB $\tau=1.17$ instead
of $\tau_{\rm eq}=4/5$ and for RF $\tau=1.59$ instead of $\tau_{\rm eq}=1$).
The anomaly observed in the exponent $\tau$ is the first fingerprint
of a discrepancy between creep events distributions and other type of
avalanches, as the depinning ones, going well beyond the anticipated
differences of critical exponents.
In \textbf{Figure \ref{fig:patterns}} it is shown
that the typical sequence of avalanches is randomly located
in space while the creep events are organized in spatio temporal
patterns very similar to earthquakes: the large events are the main
shocks that are followed by a cascade of small activated events.
The events in the cascade are the analogous of the
aftershocks which are responsible of an excess of small events
in the Gutenberg-Richter exponent as reported also in the
analysis of the real earthquakes \cite{Scholz_2002,AGGL_2016,JaglaPRL2014}
\footnote{The Gutenberg-Richter exponent $b=\frac32(\tau-1)$ for the earthquake
magnitude distribution should be smaller than the mean field prediction $3/4$,
but from seismic records one gets~\cite{JaglaPRL2014,Scholz_2002} $b\simeq 1$}.
Similar patterns for the elementary activated were observed below but near the depinning threshold~\cite{Purrello2017}.

\iffigures
\begin{figure}[t]
  \includegraphics[width=0.9\textwidth]{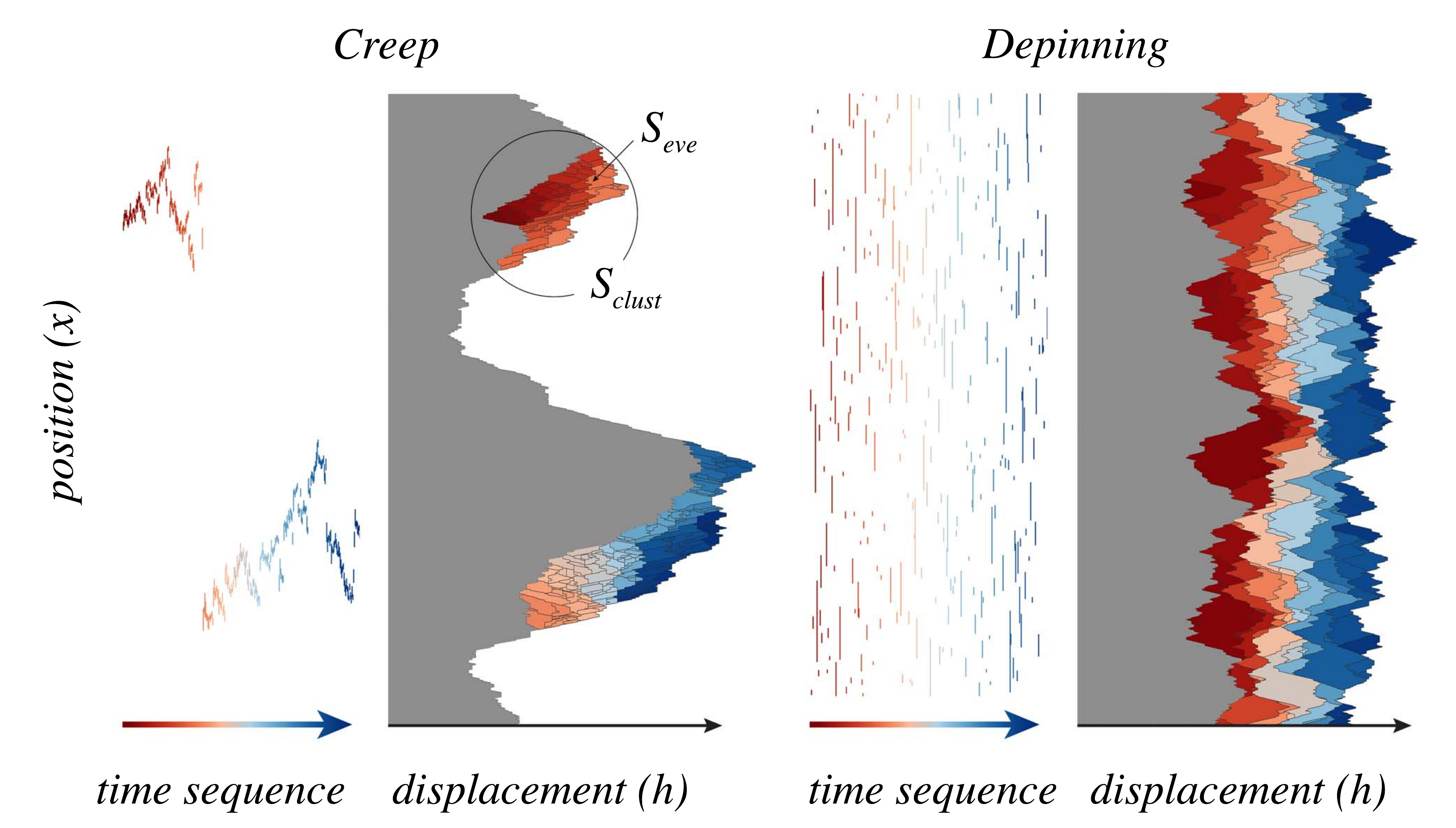}
  \caption{
  \textit{Left:} Sequence of activated events events in the creep regime.
  First, in the activity map, each segment corresponds to an event and displays its longitudinal
  length.
  The full configurations of 300 consecutive metastable states are shown immediately after.
  An individual event of size $S_{eve}$ and a cluster of size $S_{clust}$ are exemplified.
  \textit{Right:} Sequence of deterministic avalanches close to the depinning
  that appear randomly distributed in space.
  Again, both activity map and sequence of configurations are shown.
  Adapted from \cite{FerreroPRL2017}.
  }
\label{fig:patterns}
\end{figure}
\fi


In order to analyze the spatio-temporal patterns one can study the
clusters of correlated events, defined by the activated events enclosed
by a circle in  \textbf{Figure \ref{fig:patterns}}.
All details in the definition of the clusters are found in \cite{FerreroPRL2017}.

\iffigures
\begin{figure}[b]
  \begin{minipage}{.55\textwidth}
  \includegraphics[width=\textwidth]{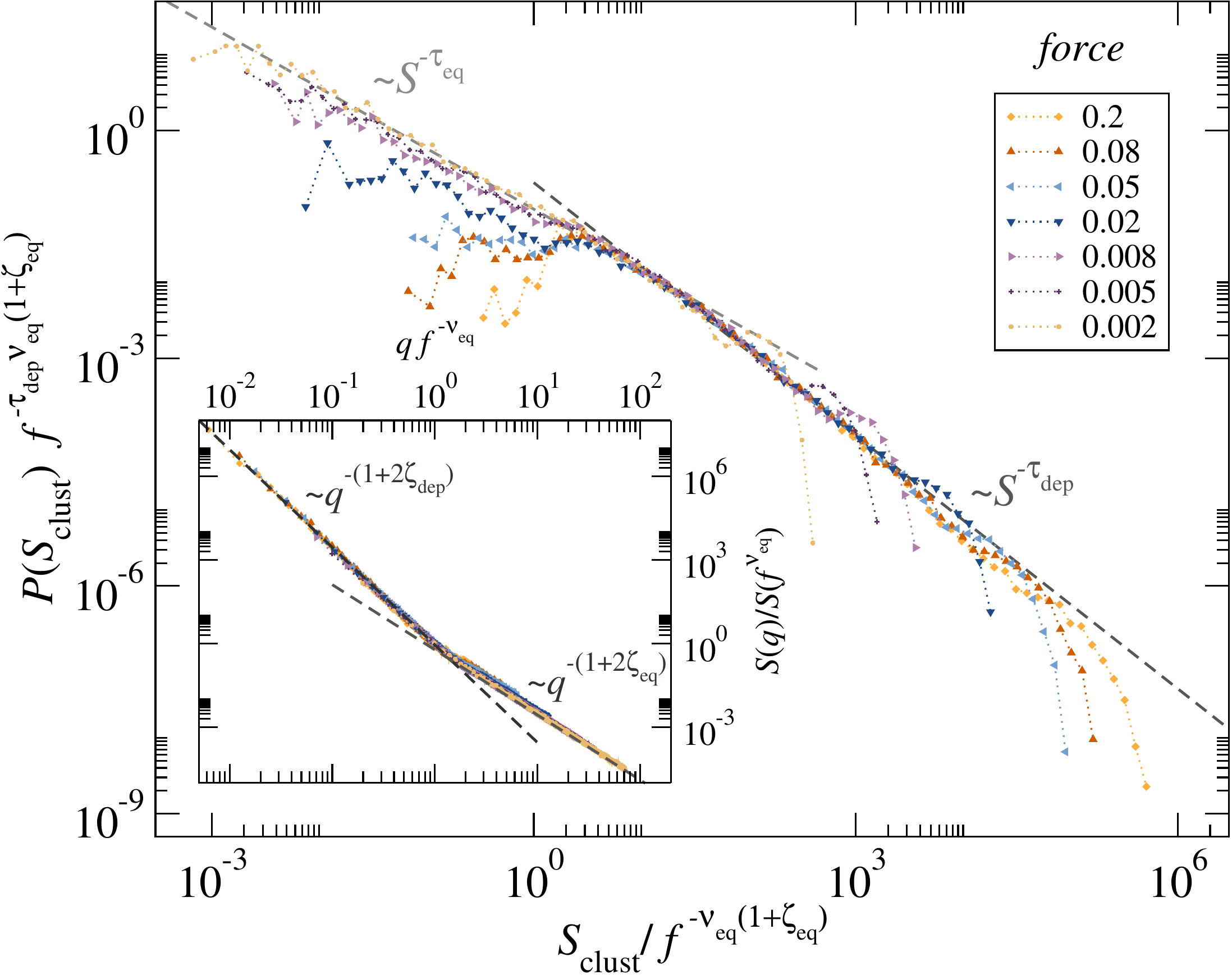}
  \end{minipage}
  \begin{minipage}{.45\textwidth}
  \includegraphics[width=\textwidth]{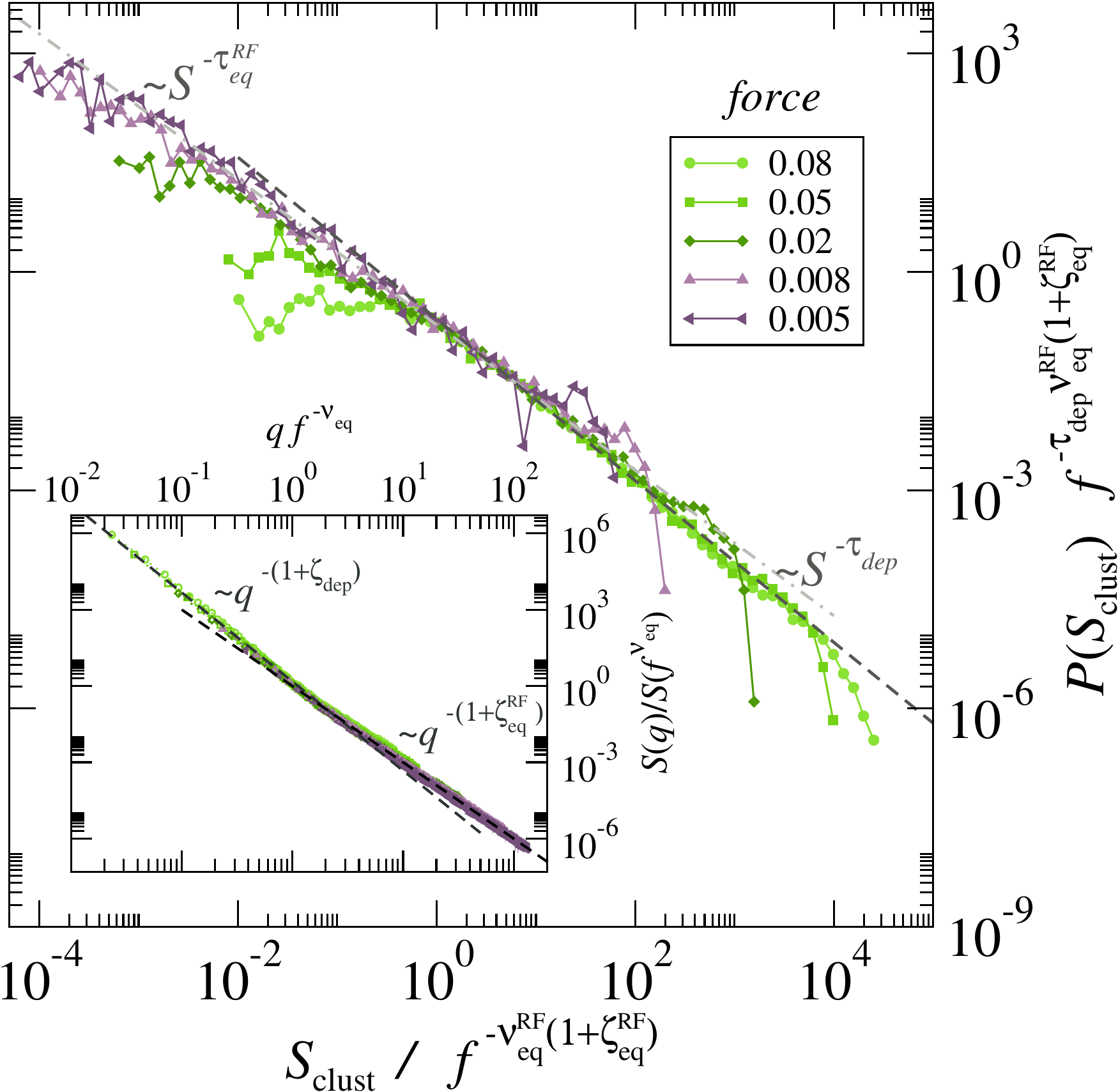}
  \end{minipage}
  \caption{Cluster area distribution $P(S_{\rm clus})$ for different forces
  for RB \textit{(left)} and RF \textit{(right)} disorder.
  A characteristic size $S_c(f)$ separates small clusters that follow \textit{equilibrium}-like
  statistics from big clusters that follow a \textit{depinning}-like one.
  This result is confirmed by the study of the
  rescaled structure factor $S(q)$ for the same forces (insets):
  a geometrical crossover is observed from \textit{equilibrium}-like
  roughness at small scales to a \textit{depinning}-like roughness at large scales.
  Adapted from \cite{FerreroPRL2017}.
  } \label{fig:clusters}
\end{figure}
\fi

Surprisingly, for both RB and RF disorder, the statistics of the clusters appear
as the one of the depinning avalanches with $\tau_{\rm dep}=1.11$ and the cut-off
controlled by the system size and diverging in the thermodynamic limit \cite{RLDW09}
(see \textbf{Figure \ref{fig:clusters}}).

\subsection{Geometry of the interface}
\label{sec:geometryoftheinterface}

An independent and complementary confirmation of these results
comes from the study of the roughness of the interface at different scales as
introduced in \cite{KoltonPRB2009}.
In practice one measures the structure factor $S(q)=\overline{u(q) u(-q)}\sim q^{-(d+2\zeta)}$
where $u(q)$ is the Fourier transform of the position of the interface
and the overline represents the average over many configurations.
The insets of \textbf{Figure \ref{fig:clusters}} shows that there exists
a crossover $1/q_c \sim \Lopt$ between two different behavior
of the roughness: at small length scales the interface seems to be at
equilibrium, while at large length scales it appears at depinning.
This observation supports the idea that the clusters are
depinning-like above a scale $\Lopt$.
Although such a result is consistent with the
predictions obtained by FRG in \cite{chauve2000},
it should be stressed that these clusters with depinning
statistics above $\Lopt$ are formed by several \textit{activated} events
rather than generated by a single deterministic move.

The coarse grained dynamics studied here is in the limit of vanishing temperature.
At finite temperature the velocity is non-zero and this induces
that the fast flow roughness becomes relevant at the large length scales
(see \textbf{Figure \ref{fig:lengths}}).
The crossover occurs at a scale $\xi$ that diverges at vanishing temperature.
The FRG proposes a scaling form for $\xi$ at low temperature and
force which depends on $f$ and $T$~\cite{chauve2000},
but this form was never tested in numerical simulation or experiments.

\iffigures
\begin{figure}[b]
  \includegraphics[width=0.8\textwidth]{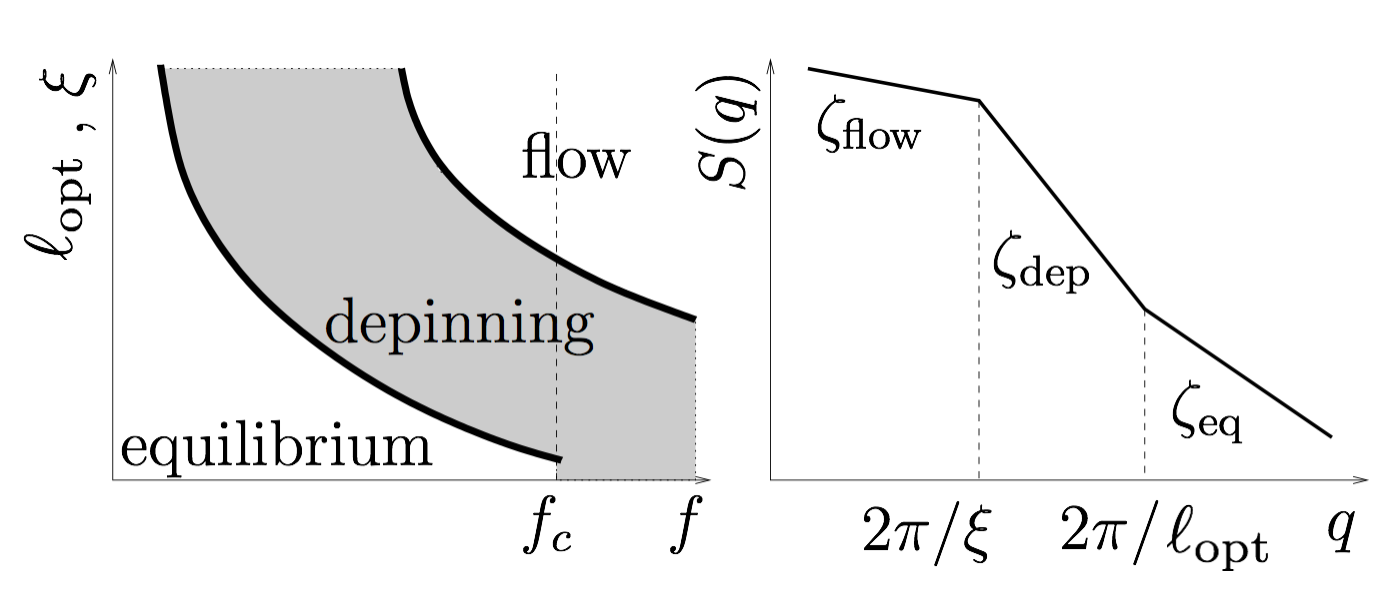}
  \caption{
  \textit{Left:} Dynamical phase diagram proposed in \cite{chauve2000}
  at finite temperature.
  Below $f_c$ the crossover between equilibrium and depinning occurs
  at the scale $\ell_{\rm opt}$.
  At finite temperature there is also a crossover at a length scale $\xi$
  between depinning and
  fast flow. However $\xi$ diverges in the limit of small temperature.
  \textit{Right:} Behavior of the roughness measured from the structure
  factor consistent with the dynamical phase diagram.
  Adapted from \cite{KoltonPRB2009}.
  } \label{fig:lengths}
\end{figure}
\fi

\subsubsection*{Quenched Edwards-Wilkinson (qEW) to quenched KPZ (qKPZ) crossover.}
\label{sec:hardconstraint}

The roughness exponent measured at large scales $\zetadep \approx 1.25$
(see the inset of \textbf{Figure~\ref{fig:clusters}}) is in agreement with
the depinning exponent of the quenched Edwards-Wilkinson
universality class.

The qEW depininning exponents are expected when the elastic interactions
are harmonic and short range as in Eq.~\ref{eq:E}.
When the interactions are anharmonic ~\cite{RossoPRL2001,Purrello2019} or a metric constraint as
Eq.~\ref{eq:hardconstraint} is present, the depinning is in the quenched KPZ universality class.
In particular the roughness exponent is expected to be $\zetadepanh\approx 0.63$~\cite{RossoPRL2001,KoltonPRB2009}.
The reasons of why simulations deep in the creep regime
(but with the metric constraint of (\ref{eq:hardconstraint}))
apparently display a crossover from $\zetaeq$ to $\zetadep$
instead of a crossover from $\zetaeq$ to $\zetadepanh$ are
analyzed in \cite{creep-long-version}.
The exponents of the qEW universality class show up at an intermediate
regime, but at very large scales the qKPZ exponents are recovered, as expected.
The crossover between the two depinning regimes is estimated to be
\begin{eqnarray}\label{eq:LanhasfunctionofLopt}
  \Lanh \propto \Lopt^{\frac{\zetadep-\zetaeq}{\zetadep-1}} .
\end{eqnarray}
For small forces the crossover occurs at very large
sizes and it cannot be observed numerically.
However, at larger forces the crossover can be observed as shown in
 \textbf{Figure \ref{fig:anharmoniccrossoverSqandPSclust2} left}
 for the structure factor and in
\textbf{Figure \ref{fig:anharmoniccrossoverSqandPSclust2} right} for the cluster
size statistics.

\iffigures
\begin{figure}
  \centering
  \begin{minipage}{.49\textwidth}
  \includegraphics[width=\textwidth]{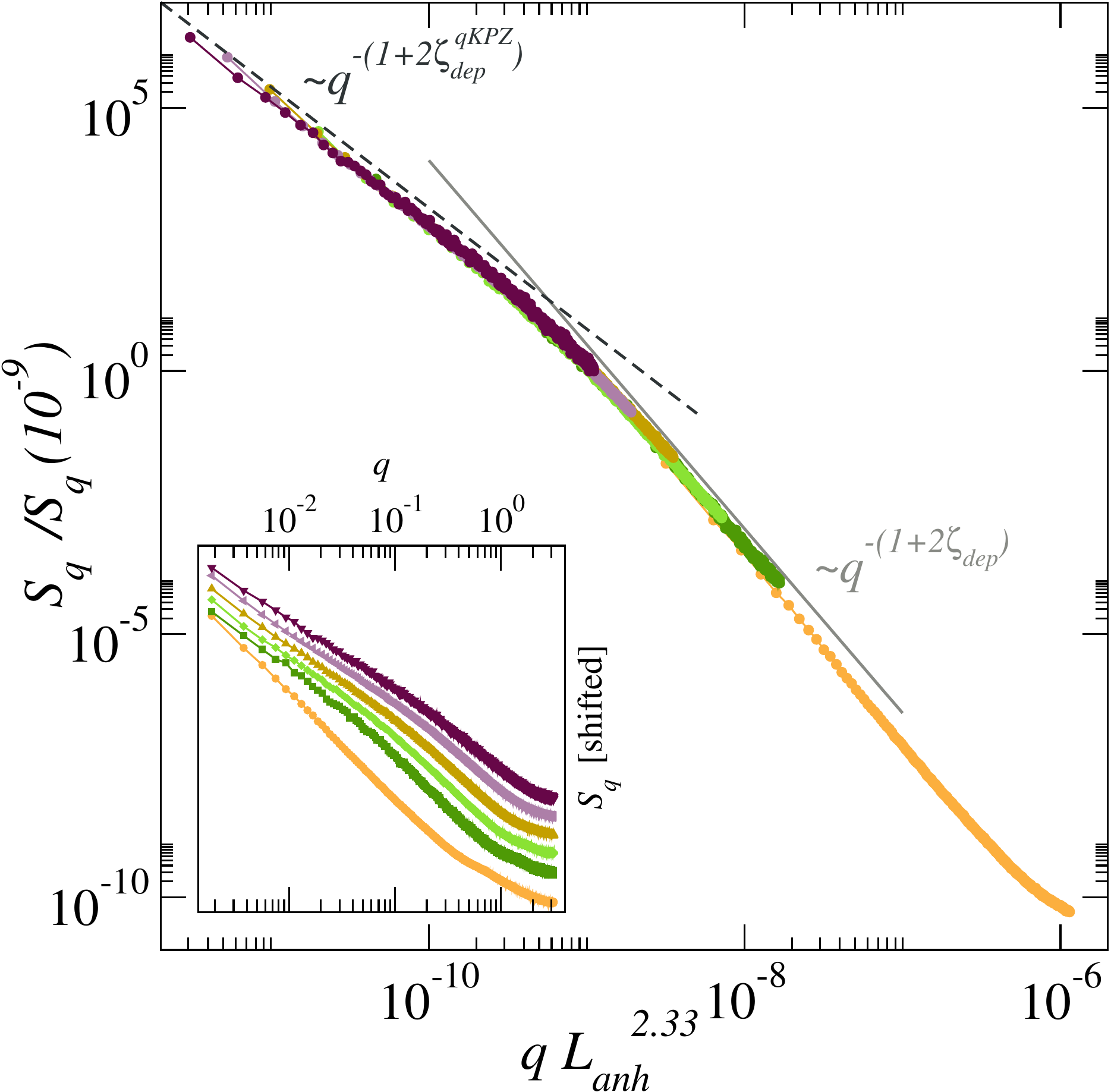}
  \end{minipage}
  \begin{minipage}{.49\textwidth}
  \includegraphics[width=\textwidth]{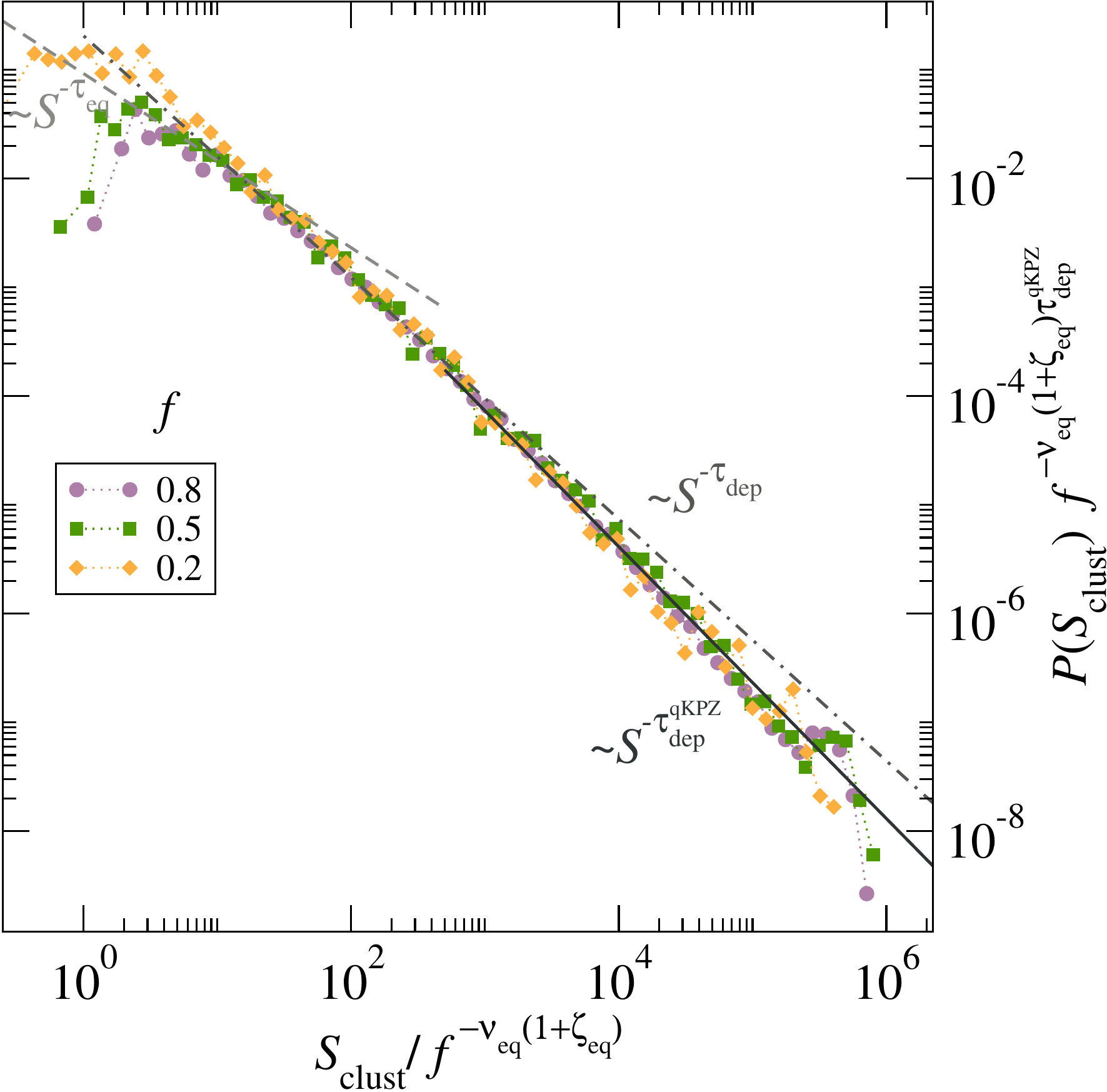}
  \end{minipage}
  \caption{
  \textit{Left:} Structure factor for the Random Bond case showing the characteristic
  lengthscale $\Lanh$ which separate the harmonic depinning regime with roughness
  exponent $\zetadep$ from the anharmonic depinning regime with exponent $\zetadepanh$,
  for different high forces $f\in\{0.2,0.5,0.6,0.7,0.8,0.9\}$, $L=3360$.
  The bottom-left inset shows the raw structure factor arbitrarily shifted in the vertical
  direction for different forces for a better display.
  The main panel shows the structure factor rescaled with $\Lanh \propto (\Lopt/\Lc)^{7/3}$,
  as proposed in Eq. \ref{eq:LanhasfunctionofLopt} for RB disorder.
  Straight gray lines are a guide to the eye, showing slopes corresponding to
  $\zetadep \simeq 1.25$ (full line) and $\zetadepanh \simeq 0.65$ (dash line).
  \textit{Right:} Cluster size distributions for $L=3360$ and $f\in\{0.2,0.5,0.8\}$.
  The anharmonic crossover has consequences in the cluster distribution
  for large cluster sizes.
  In the depinning regime the power law decay has a crossover from a regime described
  by $\taudep \approx 1.11$ to a regime described by $\taudepanh \approx 1.25$
  indicated by the two dashed lines.
  Adapted from \cite{creep-long-version}.
  } \label{fig:anharmoniccrossoverSqandPSclust2}
 \end{figure}
\fi

\subsection{Optimal Paths and Barriers}
\label{sec:barriers}

\iffigures
\begin{figure}[h]
\begin{subfigure}[h]{0.5\textwidth}
  \includegraphics[width=\textwidth]{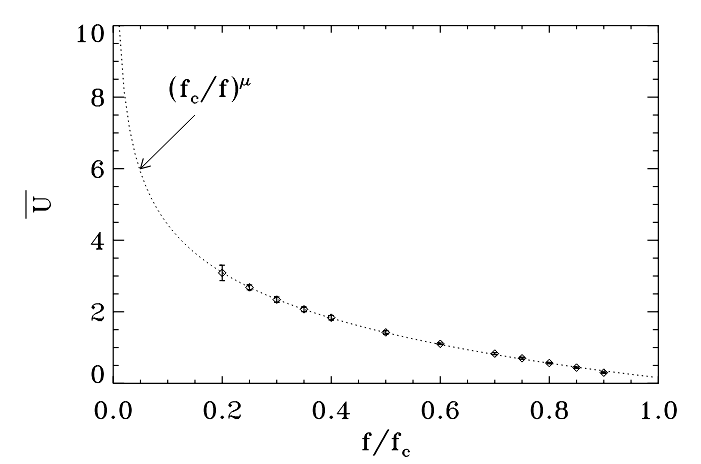}
\end{subfigure}
\begin{subfigure}[h]{0.43\textwidth}
  \includegraphics[width=\textwidth]{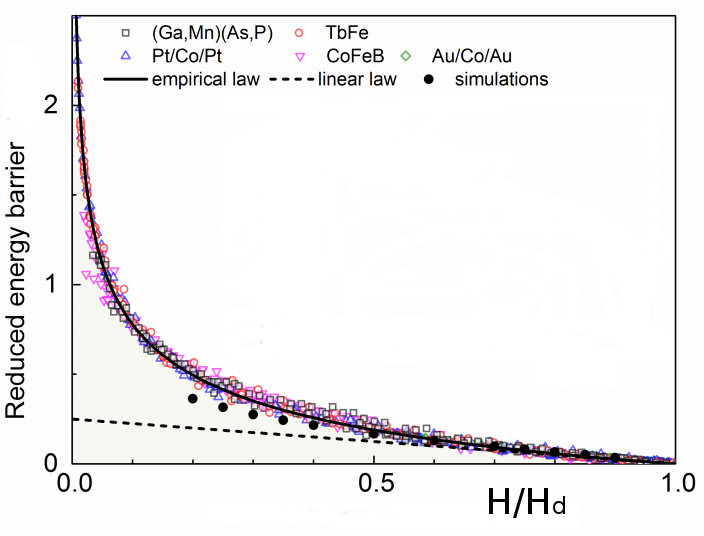}
\end{subfigure}
  \caption{
   {\textit Left:}
  Average over disorder realizations of the dominant barrier, as obtained by using the exact transition pathways algorithm.
   Adapted from \cite{KoltonPRB2009}.
 {\textit Right:} Rescaled energy barrier
 as a function of $H/H{\dep}$ for different materials and temperatures
  ranging from 10 to 315 K (25 curves in total), from~\cite{GorchonPRL2014}.
  Black circles correspond to the barriers shown on the left.}\label{fig:barrier}
\end{figure}
\fi

The exact algorithm for simulating the coarse-grained dynamics below
the depinning threshold is computationally expensive
but has the  advantage that gives access to the
energy barriers of the activated motion~\cite{KoltonPRB2009}.
If the interface moves on a torus (namely, periodic boudary
conditions are assumed both in $x$ and in $u$) the dynamics reaches a stationary
state independent on the initial condition, with a finite
sequence of metastable states $\alpha_{k}$ separated by
barriers $E_{\rm p}(\alpha_{k} \to \alpha_{k+1})$
that can be computed exactly.

Barriers are important, since the Arrhenius activation formula tell us that at vanishing temperatures
the steady state forward motion of the elastic interface is fully controlled in a finite sample by the
largest barrier $U = \max_k E_p(\alpha_{k} \to \alpha_{k+1})$ encountered in the stationary sequence
of metastable states.
The dominant configuration $\alpha_{k^*}$ such that $U = E_p(\alpha_{k^*} \to \alpha_{k^*+1})$ is
the largest barrier in a given sample plays a role similar to a ground state configuration in an
equilibrium system; in the sense that its attributes tend to dominate the average properties at low
enough temperatures (compared with the gap between the first and second largest energy barriers).

In  \textbf{Figure \ref{fig:barrier} left} we show the mean value $\overline{U}$ as a function of the force.
As expected from the creep formula $\overline{U}$ grows with decreasing the force.
Unfortunately, the computational cost of applying the exact algorithm is too high to verify the
asymptotic scaling $\overline{U} \sim f^{-\mu}$
when $f \to 0$.
When $f\to f_c$, the barrier vanishes and the size of the activated event becomes of the order of the Larkin length, 
the length for which the relative displacements are of the order of the interface thickness (or the correlation length of the disorder) \cite{AgoritsasPhysB2012}. 
This matches nicely with the behavior expected for the critical configuration at $f=f_c$.
There, the barrier is zero as the configuration is marginally stable and the soft mode is localized (Anderson-like)
with a localization length that can be identified with the Larkin length~\cite{Cao2018}.
In \textbf{Figure \ref{fig:barrier} right} we show the same quantity obtained in experiments
for different ferromagnetic domain walls.

\section{Comparison with Experiments}
\label{sec:experiments}

The creep regime has been studied in different types of domain walls.
Paradigmatic examples are domain walls in thin film ferromagnets with out of plane anisotropy~\cite{FerreCR2013},
driven by an external magnetic field or by an external electric current.
In these systems, the domain walls can be directly observed by microscopy techniques based on magneto-optic Kerr effect (MOKE).
This allows to measure the mean velocity as a function of the applied field and
the domain wall geometry.
More recently, the analysis of the images has allowed to identify
the sequence of events connecting different metastable domain wall configurations in presence of a uniform weak drive.
In this section we briefly review part of such experimental literature.
For a dedicated review of the experimental literature on magnetic domain walls
up to 2013, including reports of different values of $\mu$ and strong pinning issues, see \cite{FerreCR2013}.
As a side remark we also mention the possibility to study the creep regime of domain walls in ferroelectric materials
driven by an external electric field and observed with piezoforce microscopy~\cite{paruch2013nanoscale,kleemann2007universal}.

\subsection{Creep Velocity}

The creep law Eq.~\ref{eq:creep} was first experimentally tested in thin
ferromagnetic films (Pt/Co(0.5nm)/Pt)
driven by a magnetic field $H$ by Lemerle \textit{et al.}~\cite{LemerlePRL1998}.
They observed a clear stretched exponential behavior ($\log v \propto -H^{-\mu}$)
of the stationary mean velocity as a function of the applied field.
Rather strikingly, such law can span several decades of velocity, from almost walking speeds to the speed of nails growth.
The creep exponent $\mu$ was found to be compatible with the prediction
$\mu=(2\zetaeq-1)/(2-\zetaeq)=1/4$
where the equilibrium roughness $\zetaeq=2/3$ corresponds to a RB disorder.
A confirmation of the validity of the creep predictions was reported
later in a study of Ta/Pt/Co$_{90}$Fe$_{10}$(0.3nm)/Pt ferromagnetic thin film wires \cite{KimNature2009}.
In this paper not only Eq.~\ref{eq:creep} with $\mu \approx 1/4$ was verified,
but it was also observed a dimensional crossover ($d:1 \to 0$) in the
velocity force characteristic at low field.
Indeed, decreasing the magnetic field the length scale
$\Lopt$ grows as $\sim H^{-\nueq}$ with $\nueq = 1/(2-\zetaeq)$
up to the size of the wire's width where it saturates.
As a consequence the barrier $E_{{\rm p}} \sim \Lopt^{\theta}$ saturates inducing the breakdown of the
creep law of Eq.~\ref{eq:creep} when $\Lopt$ becomes of the order of the wire width.
A dimensional crossover ($d:1 \to 0$)
then takes place, from creep, Eq.~\ref{eq:creep}, to a TAFF like regime, Eq.~\ref{eq:TAFF}.

From the creep theory perspective the experiments
of Refs.~\cite{LemerlePRL1998,KimNature2009} hence provide crucial
information:
(\textit{i})
Although domain walls are actually two dimensional objects in three dimensional materials,
they effectively behave as a simpler one
dimensional elastic object.
In other words, the thickness of the magnetic film is smaller than $\Lopt$
and the dynamics is governed by energy barriers with $\theta(d=1)$.
(\textit{ii})
Dipolar interactions originated by stray magnetic fields
seem to be unimportant otherwise the nonlocal elasticity
would change the exponent $\mu$.
(\textit{iii})
The disorder is of RB type as for RF disorder one expects
$\zetaeq=1$, yielding $\mu=1$.
This is particularly relevant, since the nature of the DW pinning is one of the
less controlled properties of the hosting materials.

In particular since the pioneer work by Lemerle \textit{et al.} \cite{LemerlePRL1998} there have been several recent
works in thin magnetic systems reporting a consistent creep behavior
with a mean domain wall velocity showing a stretched exponential law with $\mu = 1/4$
at low enough driving fields \cite{FerreCR2013,GorchonPRL2014,JeudyPRL2016,
Pardo2017,Caballero2017,Jeudy2018,GrassiPRB2018,
Herrera2018,Domenichini2019,Shahbazi2019} and for different temperatures  \cite{GorchonPRL2014}.
The energy barrier encountered by the wall has been estimated
using the Arrhenius formula $U= - K_B T \log v/v_0$ with $v_0$
is a characteristic field independent velocity~\cite{JeudyPRL2016}.
Its behavior as a function of $H$ was found to be universal for a large family of materials:
$U$ diverges at small fields as predicted by the creep law, $U \sim H^{-\mu}$ and
vanishes at the depinning field as $U \sim (H-H_d)$ (see  \textbf{Figure \ref{fig:barrier} right}).
Both asymptotic behaviors are well described by the matching expression $U \sim (1-(H_d/H)^{\mu})$.
Moreover, the behavior experimentally observed for $U$ as
a function of $H$ is in perfect agreement
with the value $\overline{U}$ found in \cite{KoltonPRB2009}
and shown in  \textbf{Figure \ref{fig:barrier} left}.

\iffigures
\begin{figure}[h]
\begin{subfigure}[h]{0.5\textwidth}
  \includegraphics[width=\textwidth]{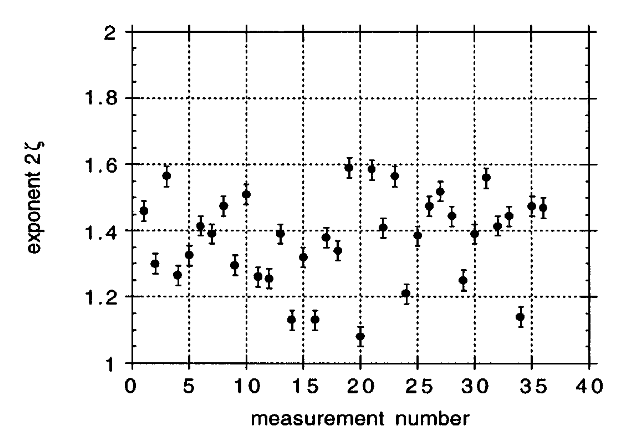}
\end{subfigure}
\begin{subfigure}[h]{0.43\textwidth}
  \includegraphics[width=\textwidth]{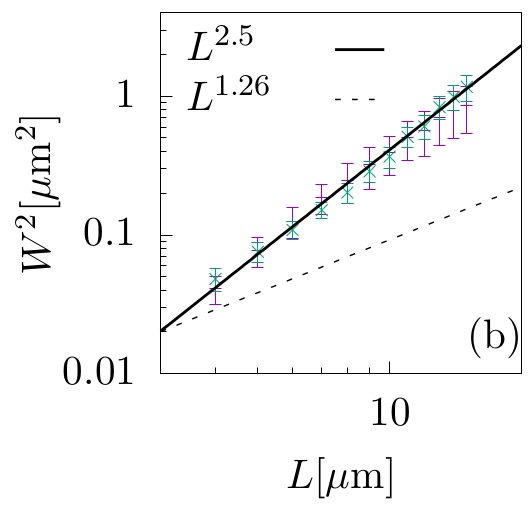}
\end{subfigure}
  \caption{\label{fig:creep-exp-images}
  \textit{Left:} Roughness exponents obtained in \cite{LemerlePRL1998} by fitting the displacement correlator
  function $\overline{ [u(x+L)-u(x)]^2 }\sim L^{2\zeta}$ with $1 \, \mu$m  $< L < 15 \, \mu$m and $v=7 \,$nm/s.
  The average exponent is $\zeta \approx 0.69 \pm 0.07$.
  \textit{Right:}
  Roughness exponent obtained in \cite{GrassiPRB2018} by fitting the detrended width.
  Different symbols correspond to two domain wall configurations at $v \approx 2\,$nm/s.
  The solid line indicates a qEW scaling $2\zetadep \approx 2.5$, the dashed line a qKPZ scaling $2\zetadepanh = 1.26$.
  }
\end{figure}
\fi

\subsection{The Roughness puzzle}

Another important test of the creep theory is to study the steady-state
roughness of the interface.
From  \textbf{Figure \ref{fig:lengths}} we expect that the width
 of a domain wall of size $L$, $w(L)$ (see Eq.~\ref{eq_roughness})
 should scale as
 \begin{equation}
 w^2(L) \sim
 \begin{cases}
 \, L^{2\zetaeq} \qquad\qquad \text{if $L<\Lopt$} \\
 \, L^{2\zetadep} \qquad\qquad \text{if $\Lopt < L < \xi$} \\
\,   L^{2 \zeta_{\rm flow}} \qquad\qquad \text{if $\xi < L$} \ .
 \end{cases}
 \end{equation}
Lemerle \textit{et al.} \cite{LemerlePRL1998} and various following works report
$\zeta \approx 0.7 \pm 0.1$, in agreement with the equilibrium value
$\zetaeq=2/3$ but far from the depinning qEW universality class $\zetadep=1.25$.
As we discuss below however, in the light of the current theory for creep and more recent experiments,
the identification of the observed $\zeta$ with $\zetaeq=2/3$ can not be justified, calling for a new reinterpretation of the data.

Recently, Gorchon et al.~\cite{GorchonPRL2014} studied field-driven domain walls in the
prototypical ultrathin Pt/Co(0.45nm)/Pt ferromagnetic films. 
By fitting the velocity force characteristics in the creep and depinning regimes,
they 
 determined the
critical depinning field $H\dep \approx 1000\;$Oe and a characteristic energy
scale $T\dep \approx 2000\;K$ at room temperature ($T=300\;K$).
%
With these values it is possible to estimate $\Lopt$ using the assumptions
of weak pinning~\cite{larkin1979,
Nattermann1990,
Demery2014}:
\begin{equation}\label{Lopt_hopt}
\begin{array}{c}
  \Lopt =   \Lc (H\dep/H)^{\nueq} \\ \\
  L_c = (k_B T\dep)/(M_s H\dep  w_c \delta )
\end{array}
\end{equation}
The microscopic Larkin length $L_c$ can be evaluated as
a function of the domain wall width $w_c$, the thickness of the sample $\delta$ and
the saturation magnetization $M_s$.
All these micromagnetic parameters are known, yielding $L_c \approx 0.04\; \mu m$
(see \cite{Jeudy2018} for the analysis for different materials).
Using a spatial resolution of $1\;\mu$m,  typical for MOKE setups
and the measured $H_{dep} \approx 1000\;$Oe one can get the condition
$H\lesssim 0.4 \;$Oe at room temperature to resolve the typical thermal
nucleous size, i.e. $\Lopt > 1 \;\mu$m.
Interestingly, $\Lopt$ was estimated in Ta/Pt/Co$_{90}$Fe$_{10}$(0.3nm)/Pt
wires~\cite{KimNature2009} with a completely different method,
observing finite size effects as the wire width $w$ was reduced.
A good scaling $\Lopt\sim H^{-\nueq}$ with 1-$d$ RB exponents,
compatible with $\zetaeq=2/3$, was found.
For these samples a field of $H=16\;$Oe gives $\Lopt \approx 0.16\;\mu$m,
remarkably in good agreement with the above estimate for the Pt/Co/Pt film.
Unfortunately, no direct roughness exponent measuremnts were reported in Ref~\cite{KimNature2009}.
%
The above estimates suggest that the range of length scales used to fit experimentally
the roughness exponent exceed  the size of $\Lopt$.
This implies that the value $\zeta \sim 0.6-0.7$ recorded in \cite{LemerlePRL1998,Shibauchi2001,Moon2013,Domenichini2019,pardo2019common}
can not be interpreted as an equilibrium exponent and
must actually correspond to the depinning regime or to the fast flow regime of roughness
(see \textbf{Figure \ref{fig:phase_diagram}})

The fast flow exponent predicted for RB or RF systems is $\zeta_{\rm flow} = 1/2$
both for RB or RF systems, quite far from the observed values.
For short range elasticity there are two universality classes at the depinning transition:
the qEW with a roughness exponent $\zetadep\simeq1.25$ and the quenched KPZ
with $\zetadepanh\simeq0.63$.
The first value is consistent with the roughness exponent obtained in \cite{GrassiPRB2018}
at low velocity, while the last value is  remarkably close to the values
at higher velocity reported in \cite{LemerlePRL1998}.
A possible way to solve this puzzle is to invoke a crossover qEW / quenched KPZ
already observed in the numerical simulations in Section~\ref{sec:hardconstraint}.
There, at low drive, the crossover occurs at very large length scales,
and the qEW exponents are measured.
At higher drive the quenched KPZ is recovered already at short distances.
To invoke such an identification however, we have to justify the presence of
a KPZ term in the effective DW equation of motion.
At least two mechanisms can justify the presence of a non-linear KPZ term:
\textit{(i)} A kinetic mechanism yields $\lambda \sim v$~\cite{kardar1986dynamic}
for interfaces driven by a pressure (i.e. driven by a force locally normal to the interface).
\textit{(ii)} A quenched disorder mechanism induced by the anisotropy of the disorder \cite{Tang1995}
or anharmonicities in the elasticity \cite{RossoPRL2001,RossoPRE2003,Purrello2019}
yields a velocity independent $\lambda$.
At the depinning transition only the second mechanism is relevant but at
the moment we lack a microscopic derivation and the presence of
crossovers between qEW and qKPZ is still under debate. 

To shed light on this puzzle another important ingredient
that should potentially be taken into account is the presence of defects such as bubbles and overhangs, at short lengthscales.
The effects of these defects on the large scale properties of the domain wall
are not yet well understood.
Large scale simulations on the 3-$d$ random field Ising model showed an anomalous behavior
of the roughness of the interface which doesn't match with the qEW prediction~\cite{clemmer2019anisotropic}
(see also~\cite{ZhouPRE2014}).

\subsection{Creep avalanches}

\iffigures
\begin{figure}[h]
\begin{subfigure}[h]{0.48\textwidth}
  \includegraphics[width=\textwidth]{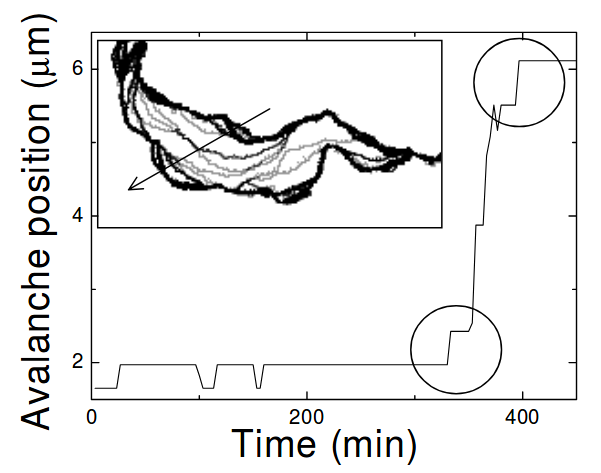}
\end{subfigure}
\begin{subfigure}[h]{0.42\textwidth}
  \includegraphics[width=\textwidth]{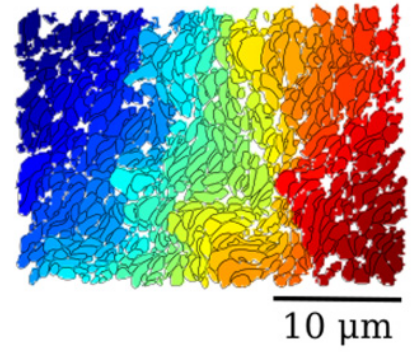}
\end{subfigure}
  \caption{\label{fig:creepevents}
\textit{Left:} Large reorganizations as obtained by Repain \textit{et al.}~\cite{RepainEPL2004} in irradiated Pt/Co/Pt thin films.
The inset shows the successive domain wall configurations in a $92 \times 28\;\mu$m$^2$ field of view.
Time interval between two images is $\Delta t=200$ s.
\textit{Right:} Sequences  of  magnetization reversal areas detected deep in the creep regime of Pt/Co/Pt thin films,
as obtained by Grassi \textit{et al.}~\cite{GrassiPRB2018}.
In this image time windows of $\Delta t=15$ s were used.
}
\end{figure}
\fi

A direct experimental access to the thermal activated events  and clusters would constitute a strong
test for the current theoretical picture.

Repain \textit{et al.} in \cite{RepainEPL2004} observed reorganizations in the creep regime
whose characteristic size qualitatively increases when lowering the field.
It is not clear if these reorganizations can be identified with the thermal activated events
as they look like chains of concatenated arcs (see inset in Fig.\ref{fig:creepevents})
suggesting the presence of strong diluted pinning.
More recently, Grassi \textit{et al.} \cite{GrassiPRB2018} performed a detailed and more quantitative
analysis in non-irradiated Pt/Co/Pt films, focusing on regions of the sample where strong pinning was not present.
They observed almost independent thermally activated reorganizations.
Their observations are consistent with the existence of ``creep avalanches''
with broad size and waiting-time distributions.
It is tempting to identify them with the clusters found in numerical simulations
discussed in Section \ref{sec:statisticsofevents}.

The quantitative experimental study of creep events remains a big experimental challenge.
The single thermally activated event or ``elementary creep event'' of Sec~\ref{sec:numericalcreep}
appears to be systematically too small to be resolved by Kerr microscopy, even for velocities of order of $v\sim 1 \, $nm/s.
Partially developed clusters appear to be accessible however, yielding indirect information
about the elementary events that control the mean creep velocity.
Understanding the effect of strong diluted pinning mixed with weak dense pinning is of
crucial importance for a quantitative analysis, since elementary activated events could
be equally associated to the collective rearrangements of typical size $\Lopt$ or to
activated depinning from strong centers.

\section{Conclusions and Perspectives}
\label{sec_Concl}

Elastic interfaces driven in disordered media represent a dramatic simplification
of physical systems, such as magnetic domain walls in disordered ferromagnets.
However, by encompassing the key interplay between elasticity and disorder, these
models are able to predict with extraordinary precision some properties
which are practically impossible to infer from more realistic microscopic approaches.
An important example is provided by the creep regime.
The theoretical picture is now well understood:
\begin{itemize}
\item The velocity versus the force characteristics displays a stretched exponential
behavior.
\item The geometrical properties of the interface show a crossover
from an equilibrium-like behavior at short length scales to
a depinning-like behavior at large length scales.
\item The dynamics displays spatio-temporal patterns (``creep avalanches'')
made of many correlated activated events.
The statistical properties of these avalanches are described by the
depinning critical point.
\end{itemize}
The creep regime is relevant for many physical systems,
ranging from fracture fronts, contact lines or
ferroelectric domain walls.
The most striking confirmation comes however from the experiments in ferromagnetic films.
There, the stretched exponential behavior of the velocity is today well
established. More recently, the analysis of the MOKE images showed
the fingerprints of an avalanche creep dynamics.

Despite of the success of the elastic interface model
many important questions remain open.
First, the statistical properties of the creep avalanches are still an experimental
challenge: the elementary events are too small to be resolved with MOKE microscopy
and the spatio-temporal correlations have not been characterized.
Second, there is a mismatch between the roughness exponents
observed in numerical simulations and the ones observed experimentally.
To find a solution for this puzzle is probably one of the biggest current challenges
in the field. 
We hope these questions will motivate further research on the universal
collective dynamics of elastic interfaces in random media.

\section*{DISCLOSURE STATEMENT}
The authors are not aware of any affiliations, memberships, funding, or financial holdings that might be perceived as affecting the objectivity of this review.

\section*{ACKNOWLEDGMENTS}
We warmly acknowledge collaborations and uncountable vivid discussions with
E. Agoritsas, S. Bustingorry, J. Curiale, G. Durin, E .A. Jagla, V. Jeudy, W. Krauth,
V. Lecomte, P. Le Doussal, P. Paruch and K. Wiese.
We acknowledge the France-Argentina project ECOS-Sud No. A16E01.
ABK acknowledges partial support from grants PICT2016-0069/FONCyT from Argentina.
EEF acknowledges support from grant PICT 2017-1202, ANPCyT (Argentina). TG support from the Swiss National Science foundation under Division II. 
This work is supported by ``Investissements d'Avenir"
LabEx PALM (ANR-10-LABX-0039-PALM) (EquiDystant project, L. Foini).

%
%
%
%


\begin{thebibliography}{96}
\expandafter\ifx\csname natexlab\endcsname\relax\def\natexlab#1{#1}\fi

\bibitem{Barrat2003}
Eds:~Barrat JL, Dalibard J, Feigelman M, Kurchan J. 2003.
Slow relaxations and nonequilibrium dynamics in condensed matter.
Springer, Berlin

\bibitem{Berthier2011}
Berthier L, Biroli G. 2011.
\textit{Reviews of Modern Physics} 83:587

\bibitem{anderson1958absence}
Anderson PW. 1958.
\textit{Physical review} 109:1492

\bibitem{evers2008anderson}
Evers F, Mirlin AD. 2008.
\textit{Reviews of Modern Physics} 80:1355

\bibitem{sethna2001crackling}
Sethna JP, Dahmen KA, Myers CR. 2001.
\textit{Nature} 410:242

\bibitem{baret2002extremal}
Baret JC, Vandembroucq D, Roux S. 2002.
\textit{Physical review letters} 89:195506

\bibitem{lin2014scaling}
Lin J, Lerner E, Rosso A, Wyart M. 2014.
\textit{Proceedings of the National Academy of Sciences} 111:14382--14387

\bibitem{nicolas2018deformation}
Nicolas A, Ferrero EE, Martens K, Barrat JL. 2018.
\textit{Reviews of Modern Physics} 90:045006

\bibitem{bonamy2011failure}
Bonamy D, Bouchaud E. 2011.
\textit{Physics Reports} 498:1--44

\bibitem{schmittbuhl1995interfacial}
Schmittbuhl J, Roux S, Vilotte JP, M{\aa}l{\o}y KJ. 1995.
\textit{Physical Review Letters} 74:1787

\bibitem{bonamy2008crackling}
Bonamy D, Santucci S, Ponson L. 2008.
\textit{Physical review letters} 101:045501

\bibitem{FerreCR2013}
Ferr{\'e} J, Metaxas PJ, Mougin A, Jamet JP, Gorchon J, Jeudy V. 2013.
\textit{Comptes Rendus Physique} 14:651 -- 666.
Disordered systems / Syst\`emes d\`esordonn\'es

\bibitem{zapperi1998dynamics}
Zapperi S, Cizeau P, Durin G, Stanley HE. 1998.
\textit{Physical Review B} 58:6353

\bibitem{paruch2013nanoscale}
Paruch P, Guyonnet J. 2013.
\textit{Comptes Rendus Physique} 14:667--684

\bibitem{kleemann2007universal}
Kleemann W. 2007.
\textit{Annu. Rev. Mater. Res.} 37:415--448

\bibitem{moulinet2004width}
Moulinet S, Rosso A, Krauth W, Rolley E. 2004.
\textit{Physical Review E} 69:035103

\bibitem{le2009height}
Le~Doussal P, Wiese KJ, Moulinet S, Rolley E. 2009.
\textit{EPL (Europhysics Letters)} 87:56001

\bibitem{dhar1999abelian}
Dhar D. 1999.
\textit{Physica A: Statistical Mechanics and its Applications} 263:4--25

\bibitem{henkel2008non}
Henkel M, Hinrichsen H, L{\"u}beck S, Pleimling M. 2008.
Non-equilibrium phase transitions.
vol.~1.
Springer

\bibitem{narayan1993}
Narayan O, Fisher DS. 1993.
\textit{Phys. Rev. B} 48:7030--7042

\bibitem{nattermann1992}
{Thomas Nattermann}, {Semjon Stepanow}, {Lei-Han Tang}, {Heiko Leschhorn}.
  1992.
\textit{J. Phys. II France} 2:1483--1488

\bibitem{Fisher1998}
{Fisher} DS. 1998.
\textit{Physics Reports} 301:113--150

\bibitem{MullerPRB2001}
M\"uller M, Gorokhov DA, Blatter G. 2001.
\textit{Phys. Rev. B} 63:184305

\bibitem{AgoritsasPhysB2012}
Agoritsas E, Lecomte V, Giamarchi T. 2012.
\textit{Physica B: Condensed Matter} 407:1725 -- 1733.
Proceedings of the International Workshop on Electronic Crystals (ECRYS-2011)

\bibitem{FerreroCR2013}
Ferrero EE, Bustingorry S, Kolton AB, Rosso A. 2013.
\textit{Comptes Rendus Physique} 14:641 -- 650.
Disordered systems / Syst\`emes d\'esordonn\'es

\bibitem{Kardar1998}
{Kardar} M. 1998.
\textit{Physics Reports} 301:85--112

\bibitem{rosso2009avalanche}
Rosso A, Le~Doussal P, Wiese KJ. 2009{\natexlab{a}}.
\textit{Physical Review B} 80:144204

\bibitem{kolton2019distribution}
Kolton AB, Doussal PL, Wiese KJ. 2019.
\textit{{EPL} (Europhysics Letters)} 127:46001

\bibitem{papanikolaou2011universality}
Papanikolaou S, Bohn F, Sommer RL, Durin G, Zapperi S, Sethna JP. 2011.
\textit{Nature Physics} 7:316

\bibitem{laurson2013evolution}
Laurson L, Illa X, Santucci S, Tallakstad KT, M{\aa}l{\o}y KJ, Alava MJ. 2013.
\textit{Nature communications} 4:2927

\bibitem{Scholz_2002}
Scholz CH. 2002.
The mechanics of earthquakes and faulting.
Cambridge university press

\bibitem{JaglaJGR2010}
Jagla E, Kolton A. 2010.
\textit{Journal of Geophysical Research: Solid Earth} 115

\bibitem{JaglaPRL2014}
Jagla EA, Landes FP, Rosso A. 2014.
\textit{Phys. Rev. Lett.} 112:174301

\bibitem{janicevic2016interevent}
Jani{\'c}evi{\'c} S, Laurson L, M{\aa}l{\o}y KJ, Santucci S, Alava MJ. 2016.
\textit{Physical review letters} 117:230601

\bibitem{KoltonPRL2006}
Kolton AB, Rosso A, Giamarchi T, Krauth W. 2006.
\textit{Phys. Rev. Lett.} 97:057001

\bibitem{Fedorenko2006}
Fedorenko AA, Le~Doussal P, Wiese KJ. 2006.
\textit{Phys. Rev. E} 74:061109

\bibitem{FerreroPRE2013}
Ferrero EE, Bustingorry S, Kolton AB. 2013.
\textit{Phys. Rev. E} 87:032122

\bibitem{narayan1992}
Narayan O, Fisher DS. 1992.
\textit{Phys. Rev. B} 46:11520--11549

\bibitem{ledoussal2002}
Le~Doussal P, Wiese KJ, Chauve P. 2002.
\textit{Phys. Rev. B} 66:174201

\bibitem{fisher1985}
Fisher DS. 1985.
\textit{Phys. Rev. B} 31:1396--1427

\bibitem{alessandro1990}
Alessandro B, Beatrice C, Bertotti G, Montorsi A. 1990.
\textit{Journal of Applied Physics} 68:2908--2915

\bibitem{LeDoussalPhysC1995}
Doussal PL, Vinokur VM. 1995.
\textit{Physica C: Superconductivity} 254:63 -- 68

\bibitem{Joanny1984}
Joanny J, De~Gennes PG. 1984.
\textit{The journal of chemical physics} 81:552--562

\bibitem{Gao1989}
Gao H, Rice JR. 1989.
\textit{Journal of applied mechanics} 56:828--836

\bibitem{KoltonJagla2018}
Kolton AB, Jagla EA. 2018.
\textit{Phys. Rev. E} 98:042111

\bibitem{chauve2000}
Chauve P, Giamarchi T, Le~Doussal P. 2000.
\textit{Phys. Rev. B} 62:6241--6267

\bibitem{LeDoussalPRE2013}
Le~Doussal P, Wiese KJ. 2013.
\textit{Phys. Rev. E} 88:022106

\bibitem{priol2019universal}
Priol CL, Doussal PL, Ponson L, Rosso A. 2019.
\textit{arXiv preprint arXiv:1909.09075}

\bibitem{Rosso2002}
Rosso A, Krauth W. 2002.
\textit{Phys. Rev. E} 65:025101

\bibitem{RossoPRE2003}
Rosso A, Hartmann AK, Krauth W. 2003.
\textit{Phys. Rev. E} 67:021602

\bibitem{ramanathan1998}
Ramanathan S, Fisher DS. 1998.
\textit{Phys. Rev. B} 58:6026--6046

\bibitem{leschhorn1993}
Leschhorn H. 1993.
\textit{Physica A: Statistical Mechanics and its Applications} 195:324 -- 335

\bibitem{Zoia2007}
Zoia A, Rosso A, Kardar M. 2007.
\textit{Phys. Rev. E} 76:021116

\bibitem{middleton1995}
Middleton AA. 1995.
\textit{Phys. Rev. E} 52:R3337--R3340

\bibitem{Tang1995}
Tang LH, Kardar M, Dhar D. 1995.
\textit{Phys. Rev. Lett.} 74:920--923

\bibitem{Buldyrev1993}
Buldyrev SV, Havlin S, Stanley HE. 1993.
\textit{Physica A: Statistical Mechanics and its Applications} 200:200--211

\bibitem{RossoPRL2001}
Rosso A, Krauth W. 2001{\natexlab{a}}.
\textit{Phys. Rev. Lett.} 87:187002

\bibitem{kardar1986dynamic}
Kardar M, Parisi G, Zhang YC. 1986.
\textit{Physical Review Letters} 56:889

\bibitem{Tang1992}
Tang LH, Leschhorn H. 1992.
\textit{Phys. Rev. A} 45:R8309--R8312

\bibitem{Jeong1996}
Jeong H, Kahng B, Kim D. 1996.
\textit{Phys. Rev. Lett.} 77:5094--5097

\bibitem{Atis2015}
Atis S, Dubey AK, Salin D, Talon L, Le~Doussal P, Wiese KJ. 2015.
\textit{Phys. Rev. Lett.} 114:234502

\bibitem{LemerlePRL1998}
Lemerle S, Ferr\'e J, Chappert C, Mathet V, Giamarchi T, Le~Doussal P. 1998.
\textit{Phys. Rev. Lett.} 80:849--852

\bibitem{KimNature2009}
Kim KJ, Lee JC, Ahn SM, Lee KS, Lee CW, et~al. 2009.
\textit{Nature} 458:740 EP --

\bibitem{JeudyPRL2016}
Jeudy V, Mougin A, Bustingorry S, Savero~Torres W, Gorchon J, et~al. 2016.
\textit{Phys. Rev. Lett.} 117:057201

\bibitem{Ioffe1987}
Ioffe LB, Vinokur VM. 1987.
\textit{Journal of Physics C: Solid State Physics} 20:6149--6158

\bibitem{Nattermann1987}
Nattermann T. 1987.
\textit{Europhysics Letters ({EPL})} 4:1241--1246

\bibitem{vinokur1996}
Vinokur VM, Marchetti MC, Chen LW. 1996.
\textit{Phys. Rev. Lett.} 77:1845--1848

\bibitem{Anderson1964}
Anderson PW, Kim YB. 1964.
\textit{Rev. Mod. Phys.} 36:39--43

\bibitem{KoltonPRB2009}
Kolton AB, Rosso A, Giamarchi T, Krauth W. 2009.
\textit{Phys. Rev. B} 79:184207

\bibitem{drossel1995}
Drossel B, Kardar M. 1995.
\textit{Phys. Rev. E} 52:4841--4852

\bibitem{kolton2005creep}
Kolton AB, Rosso A, Giamarchi T. 2005.
\textit{Phys. Rev. Lett.} 94:047002

\bibitem{RossoPRB2001}
Rosso A, Krauth W. 2001{\natexlab{b}}.
\textit{Physical Review B} 65:012202

\bibitem{FerreroPRL2017}
Ferrero EE, Foini L, Giamarchi T, Kolton AB, Rosso A. 2017.
\textit{Phys. Rev. Lett.} 118:147208

\bibitem{AGGL_2016}
Arcangelis L, Godano C, Grasso JR, Lippiello E. 2016.
\textit{to be published in Physics Report}

\bibitem{Purrello2017}
Purrello VH, Iguain JL, Kolton AB, Jagla EA. 2017.
\textit{Phys. Rev. E} 96:022112

\bibitem{RLDW09}
Rosso A, Le~Doussal P, Wiese KJ. 2009{\natexlab{b}}.
\textit{Physical Review B} 80:144204

\bibitem{Purrello2019}
Purrello VH, Iguain JL, Kolton AB. 2019.
\textit{Phys. Rev. E} 99:032105

\bibitem{creep-long-version}
Ferrero EE, Foini L, Giamarchi T, Kolton AB, Rosso A. 2019.
Non-linear elasticity and collective events in domain wall creep dynamics.
In preparation

\bibitem{GorchonPRL2014}
Gorchon J, Bustingorry S, Ferr\'e J, Jeudy V, Kolton AB, Giamarchi T. 2014.
\textit{Phys. Rev. Lett.} 113:027205

\bibitem{Cao2018}
Cao X, Bouzat S, Kolton AB, Rosso A. 2018.
\textit{Phys. Rev. E} 97:022118

\bibitem{Pardo2017}
Diaz~Pardo R, Savero~Torres W, Kolton AB, Bustingorry S, Jeudy V. 2017.
\textit{Phys. Rev. B} 95:184434

\bibitem{Caballero2017}
Caballero NB, Fern\'andez~Aguirre I, Albornoz LJ, Kolton AB, Rojas-S\'anchez
  JC, et~al. 2017.
\textit{Phys. Rev. B} 96:224422

\bibitem{Jeudy2018}
Jeudy V, D\'{\i}az~Pardo R, Savero~Torres W, Bustingorry S, Kolton AB. 2018.
\textit{Phys. Rev. B} 98:054406

\bibitem{GrassiPRB2018}
Grassi MP, Kolton AB, Jeudy V, Mougin A, Bustingorry S, Curiale J. 2018.
\textit{Phys. Rev. B} 98:224201

\bibitem{Herrera2018}
Herrera~Diez L, Jeudy V, Durin G, Casiraghi A, Liu YT, et~al. 2018.
\textit{Phys. Rev. B} 98:054417

\bibitem{Domenichini2019}
Domenichini P, Quinteros CP, Granada M, Collin S, George JM, et~al. 2019.
\textit{Phys. Rev. B} 99:214401

\bibitem{Shahbazi2019}
Shahbazi K, Kim JV, Nembach HT, Shaw JM, Bischof A, et~al. 2019.
\textit{Phys. Rev. B} 99:094409

\bibitem{larkin1979}
Larkin AI, Ovchinnikov YN. 1979.
\textit{Journal of Low Temperature Physics} 34:409--428

\bibitem{Nattermann1990}
Nattermann T, Shapir Y, Vilfan I. 1990.
\textit{Phys. Rev. B} 42:8577--8586

\bibitem{Demery2014}
D{\'{e}}mery V, Lecomte V, Rosso A. 2014.
\textit{Journal of Statistical Mechanics: Theory and Experiment} 2014:P03009

\bibitem{Shibauchi2001}
Shibauchi T, Krusin-Elbaum L, Vinokur VM, Argyle B, Weller D, Terris BD. 2001.
\textit{Phys. Rev. Lett.} 87:267201

\bibitem{Moon2013}
Moon KW, Kim DH, Yoo SC, Cho CG, Hwang S, et~al. 2013.
\textit{Phys. Rev. Lett.} 110:107203

\bibitem{pardo2019common}
Pardo RD, Moisan N, Albornoz L, Lemaitre A, Curiale J, Jeudy V. 2019.
Common universal behaviors of magnetic domain walls driven by spin-polarized
  electrical current and magnetic field

\bibitem{clemmer2019anisotropic}
Clemmer JT, Robbins MO. 2019.
\textit{Phys. Rev. E} 100:042121

\bibitem{ZhouPRE2014}
Zhou NJ, Zheng B. 2014.
\textit{Phys. Rev. E} 90:012104

\bibitem{RepainEPL2004}
{Repain, V.}, {Bauer, M.}, {Jamet, J.-P.}, {Ferr\'e, J.}, {Mougin, A.}, et~al.
  2004.
\textit{Europhys. Lett.} 68:460--466

\end{thebibliography}
\end{document}